\documentclass[10pt,a4paper]{article}

\usepackage{times} 
\usepackage{graphicx}            
\usepackage{amsmath}             
\usepackage{amssymb, bm}  
\usepackage{mathrsfs}          
\usepackage{booktabs}           
\usepackage{array}              
\usepackage{multirow}           
\usepackage{lineno}  
\usepackage{float}   
\usepackage{pdflscape}       
\usepackage[numbers,sort&compress]{natbib} 
\usepackage{setspace}           
\usepackage[colorlinks=true, allcolors=blue]{hyperref} 
\usepackage{titlesec} 
\selectfont 
\usepackage{titlesec}
\usepackage{etoolbox}
\usepackage{setspace}
\usepackage[left=2.cm,right=2.cm,top=2.cm,bottom=2.cm]{geometry}

\titleformat{\section}
{\fontsize{10}{9}\selectfont\bfseries\raggedright} 
{\thesection.} 
{0.3em} 
{}
\titlespacing*{\section}{0pt}{10pt}{9pt}

\titleformat{\subsection}
{\fontsize{10}{9}\selectfont\bfseries\raggedright} 
{\thesubsection.}
{0.3em}
{}
\titlespacing*{\subsection}{0pt}{10pt}{9pt}

\titleformat{\subsubsection}
{\fontsize{10}{9}\selectfont\bfseries\itshape\raggedright} 
{\thesubsubsection.}
{0.3em}
{}
\titlespacing*{\subsubsection}{0pt}{10pt}{6pt}

\title{Self-organized pattern synchronization modulated by stochasticity in coupled plankton ecosystems}
\author{
	Ju Kang\textsuperscript{1,2}, 
	Yiyuan Niu\textsuperscript{3},
	Tousheng Huang\textsuperscript{4},
	Quan-Xing Liu\textsuperscript{5},
	Chengjin Chu\textsuperscript{1,*}
}
\date{}

\newcommand{\affil}[2]{%
	\textsuperscript{#1}#2%
}
\makeatletter
\renewcommand{\maketitle}{
	\begin{center}
		{\LARGE\bfseries \@title\par}
		\vspace{1em}  
		{\normalsize \@author\par}
		\vspace{-0.3em} 
		\setlength{\parskip}{1pt}
		\@date
	\end{center}
}
\makeatother	

\begin{document}
	
	\maketitle
	
	\begin{center}
		\affil{1}{School of Ecology, Sun Yat-sen University, Shenzhen 518107, China}\\
		\affil{2}{School of Artificial Intelligent Medicine, Guilin Medical University, Guilin 541199, China}\\
		\affil{3}{School of Physics, Sun Yat-sen University, Guangzhou 510275, China}\\
		\affil{4}{Research Center for Engineering Ecology and Nonlinear Science, North China Electric Power University, Beijing 102206, China}\\
		\affil{5}{School of Mathematical Sciences, Shanghai Jiao Tong University, Shanghai 200240, China}\\    
		\affil{*}{Corresponding author: \href{mailto:chuchjin@mail.sysu.edu.cn}{chuchjin@mail.sysu.edu.cn}}\\	
	\end{center}	
	
\begin{abstract}
	Spatial patterning and synchronization are pervasive features of plankton communities, yet how such patterns persist coherently under environmental noise is not fully understood. In vertically structured aquatic ecosystems, plankton populations are often organized into distinct layers, raising the question of how interactions between layers shape both spatial self-organization and robustness. Here, we develop a spatiotemporal ecosystem model of a two-layer plankton community of non-toxic phytoplankton, toxic phytoplankton and zooplankton, and use it to examine the role of passive diffusive coupling under stochastic environmental fluctuations. We show that interlayer diffusion induces a sharp transition from independent, layer-specific Turing patterns to spatially synchronized patterns once the coupling strength exceeds a critical threshold. Importantly, the same coupling mechanism also markedly enhances the stability of spatial patterns against environmental noise, extending their persistence far beyond that of non-coupled layers. A linear-response analysis shows that the coupling damps the difference between the two layers at a rate that grows with the coupling strength. Moreover, we find a trophic hierarchy in noise sensitivity, in which zooplankton are more vulnerable to noise than either phytoplankton type. Together, these results show that passive diffusive coupling between layers both synchronizes spatial patterns and buffers them against environmental noise, and thus offers an explanation for the persistence of coherent plankton patterns in fluctuating aquatic environments.
\end{abstract}	

\section{Introduction}  
Spatial pattern formation is a pervasive and striking feature of plankton communities in aquatic ecosystems, manifesting as patchy distributions across both horizontal and vertical dimensions. Such spatial heterogeneity has been widely documented in field observations and experiments and is thought to play a central role in regulating species interactions, ecosystem functioning, and biogeochemical cycling~\cite{NatalieLGeyer2018,SimonvanGennip2016,MatiasScheinin2020,KellyL.Robinson2021,JohanvandeKoppel2005}. Understanding the mechanisms that generate and sustain these patterns has therefore long been a central objective in aquatic ecology.

Over the past decades, a broad range of biological and physical processes, including predation, migration, diffusion, turbulence, and hydrodynamic forcing, have been proposed as drivers of plankton spatial self-organization~\cite{FeifanZhang2022,ToushengHuangCNSNS,ToushengHuang2023,CaroLFolt1999,LevinSA1976,AlexanderMedvinsky2002,JenniferPrairie2012,Edward R.Abraham1998,WilliamMDurham2013,APMartin2003}. Many of these mechanisms are capable of producing self-organized spatial patterns in idealized settings, particularly within single layers or effectively isolated spatial domains. Recent experimental studies have further revealed that stochasticity intrinsic to biological systems can itself drive Turing-like pattern formation, as demonstrated in synthetic bacterial populations~\cite{Karig2018} and in the development of trichomes in \textit{Arabidopsis} leaves~\cite{Patti2023}. However, most existing studies, whether focused on ecological or molecular systems, have concentrated on pattern formation in isolation within single-layer or isolated domains, rather than on the persistence and coherence of patterns in realistic environments where multiple spatial domains are dynamically coupled and continuously exposed to environmental fluctuations.

Recent theoretical advances have clarified how the conditions for Turing instabilities are modified when several reaction-diffusion layers are coupled, and how the topology of the coupling and the strength of inter-layer diffusion select the resulting spatial structures~\cite{Asllani2014,CalderonBarreto2026}. However, these studies have largely focused on the mathematical criteria for pattern formation in idealized network structures. The question of how such coupling mechanisms interact with environmental stochasticity to determine the persistence and robustness of spatial patterns in ecologically realistic systems, particularly those with trophic structure and biologically motivated movement rules, remains less explored.

In aquatic environments, 	extcolor{blue}{zooplankton frequently exhibit selective feeding} on toxic versus non-toxic phytoplankton species~\cite{TurriffN1995,MetteSchultz2009}. Such avoidance is not always absolute; under conditions of limited resources, some zooplankton initially prefer non-toxic prey but gradually shift to toxic alternatives when necessary~\cite{TurriffN1995,MetteSchultz2009}. This flexible foraging strategy implies that prey quality, including toxicity, can at times override established selection criteria such as size. For example, \textit{Artemia salina} has been shown to prefer larger non-toxic prey over smaller toxic ones~\cite{MetteSchultz2010}. Thus, nuanced avoidance behavior can significantly influence population persistence and community structure. Toxic prey avoidance in zooplankton is shaped more strongly by behavioral adaptation than by physiological resistance~\cite{GregoryJTeegarden1999}. Although physiological tolerance to toxins may evolve over time~\cite{Starkweather1983,TakayukiHanazato1987,RollandFulton1988}, behavioral strategies such as selective ingestion often constitute the primary defense mechanism~\cite{DeMott1991,Kurmayer1999}. Phytoplankton-released toxins can serve as a defense against grazing~\cite{GregoryJTeegarden1999,AlbertCalbet2003} and may also negatively affect competing species~\cite{LESchmidt2001,AllanCembella2003}. These toxins exert a range of adverse effects on zooplankton, from deterring feeding to impairing physiological function and reducing reproductive success~\cite{GregoryJTeegarden1999,GordonVWolfe2000,Sukhanov2002}. The intensity of zooplankton avoidance behavior, which varies with the availability of non-toxic prey, is therefore a key factor influencing species coexistence, ecosystem stability, and potential harmful algal bloom formation, as indicated by modeling studies~\cite{JordiSole2006}.

Vertical structuring is a ubiquitous characteristic of aquatic systems. Stratification of the water column, together with active vertical migration of zooplankton and passive transport of phytoplankton, naturally couples plankton communities across depth layers~\cite{JenniferPrairie2012,KanchanaBandara2021,BingzhangChen2020,ShaneARichards1996,MaciejZGliwicz1986,StefanoSimoncelli2019,MonicaWinder2003,XabierIrigoien2004,RainerKiko2019}. Such coupling establishes an extended ecosystem in which biological and physical processes in one layer can influence pattern dynamics in another. Notable examples include euphausiids (e.g., \textit{Euphausia mucronata}) and the semi-pelagic squat lobster \textit{Pleuroncodes monodon} in the Peruvian upwelling system, which undertake diel vertical migrations connecting surface and deep waters, thereby actively linking biogeochemical processes across depth layers~\cite{RainerKiko2019}.
Empirical and theoretical studies have suggested that vertical coupling may promote synchronization of spatial patterns across layers~\cite{CaroLFolt1999,BingzhangChen2020,KendraLDaly1993,AndrewYuMorozov2007,ToushengHuangCNSNS,ToushengHuang2023,KaleviSalonen2024}. Despite this extensive body of work, most existing studies have focused on pattern formation within single layers or under effectively isolated spatial domains, often by isolating individual biological processes or physical drivers~\cite{CaroLFolt1999,LevinSA1976,AlexanderMedvinsky2002}. However, the mechanisms by which vertically coupled plankton systems achieve and maintain synchronized spatial self-organization remain incompletely understood~\cite{YonghongChe2006}.

In contrast to prior studies that have primarily focused on deterministic pattern formation and classification~\cite{ToushengHuangCNSNS,ToushengHuang2023}, we address a fundamental question that remains unexplored: how do vertically coupled plankton communities maintain synchronized spatial patterns in the presence of persistent environmental stochasticity? While Huang et al.~\cite{ToushengHuangCNSNS,ToushengHuang2023} established the dispersion-relation peak-merging mechanism for pattern synchronization, they did not investigate how such synchronized patterns respond to noise, nor did they examine whether coupling confers stability advantages against environmental fluctuations. Our study is therefore concerned with the interplay between diffusive coupling, stochastic forcing, and trophic structure. Furthermore, our model includes biologically motivated cross-diffusion terms (zooplankton movement in response to phytoplankton gradients), which are absent from previous coupled-layer plankton models and which fundamentally alter the pattern-forming dynamics and noise response.

This gap becomes particularly pronounced in the presence of environmental stochasticity. Aquatic ecosystems are inherently noisy, with fluctuating physical conditions such as turbulence, nutrient supply, and hydrodynamic forcing acting as persistent sources of perturbation. In reaction-diffusion systems, stochasticity is generally expected to disrupt self-organized spatial patterns. Nonetheless, coherent and synchronized plankton patterns are frequently observed in nature, even under highly variable environmental conditions. How vertically coupled plankton communities reconcile this apparent tension between noise-induced disruption and persistent spatial coherence remains an open and fundamental problem.

In this work, we address this problem by developing a spatiotemporal ecosystem model of a two-layer plankton community coupled through passive diffusive exchange. By deliberately focusing on physical coupling rather than species-specific behavioral complexity, we isolate the role of diffusion in shaping spatial dynamics under stochastic perturbations. We show that increasing coupling strength induces a sharp transition from independent, layer-specific Turing patterns to fully synchronized spatial patterns once a critical threshold is exceeded. Importantly, the same diffusive coupling mechanism substantially enhances the robustness of spatial patterns against environmental noise, allowing synchronized structures to persist far beyond the regime where non-coupled layers lose coherence. Furthermore, we find that noise sensitivity differs systematically across trophic levels, with zooplankton populations exhibiting greater vulnerability than phytoplankton. Together, these results identify passive diffusive coupling as a core mechanism that simultaneously promotes spatial synchronization and robustness, offering a mechanistic explanation for the persistence of coherent plankton patterns in fluctuating aquatic environments.

\section{Results}
\subsection{A coupled two-layer reaction-diffusion framework}
In stratified water columns, the number of coupled plankton systems is determined by the number of vertical layers. A greater number of layers leads to more intricate dynamic patterns in the coupled system, making it more challenging to decipher the underlying nonlinear self-organization mechanisms. The degree of stratification, both in terms of the number of layers and the strength of gradients, shapes the physical template upon which biological self-organization occurs. Importantly, stratification does not imply the absence of turbulence; rather, it modulates turbulent processes that may serve as sources of environmental stochasticity. These stochastic disturbances, in turn, interact with the deterministic dynamics of pattern formation, potentially influencing the stability and synchronization of spatial structures across layers. Therefore, to focus on the core mechanisms, namely Turing instability, diffusive coupling, and their interplay with environmental stochasticity, this study analyzes the simplest case of two-layer stratification, an ecologically realistic scenario where two subsystems, each comprising non-toxic phytoplankton, toxic phytoplankton, and zooplankton, are diffusively coupled. Building on this two-layer model, we analyze a vertically stratified planktonic ecosystem using coupled reaction-diffusion equations that resolve horizontal spatial structure within two vertically distinct layers. The model retains three state variables per layer: non-toxic phytoplankton $N_i(x,y,t)$, toxic phytoplankton $T_i(x,y,t)$ and zooplankton $Z_i(x,y,t)$ for layer index $i\in\{1,2\}$. In natural aquatic ecosystems, non-toxic phytoplankton ($N_i$) primarily corresponds to green algae (e.g., \textit{Chlorella}, \textit{Scenedesmus}) and diatoms, which serve as high-quality food sources for zooplankton. Toxic phytoplankton ($T_i$) corresponds to cyanobacteria (e.g., \textit{Microcystis aeruginosa}, \textit{Raphidiopsis raciborskii}), which produce microcystins that adversely affect zooplankton feeding, growth, and reproduction. Zooplankton ($Z_i$) encompasses functional groups such as copepods (e.g., \textit{Mesocyclops}, \textit{Eudiaptomus}) and cladocerans (e.g., \textit{Daphnia}, \textit{Moina}), which exhibit different feeding selectivities; copepods are capable of selective feeding based on food quality, whereas cladocerans are non-selective filter-feeders. These grazers play a pivotal role in linking primary producers to higher trophic levels and regulating phytoplankton community structure through top-down control. Each layer occupies a square domain $\Omega=[0,L]\times[0,L]\subset\mathbb{R}^2$ with smooth boundary $\partial\Omega$; homogeneous Neumann (no-flux) boundary conditions are imposed on all species (see below). The model of two coupled non-toxic phytoplankton, toxic phytoplankton and zooplankton well-mixed system can be written as follows:
	\begin{equation}	
		\begin{cases}
			\frac{\partial {{N}_{1}}\left( x,y,t \right)}{\partial t}=f_{1}({N}_{1},{T}_{1},{Z}_{1})+{{d}_{11}}\Delta {{N}_{1}}+{{h}_{11}}\left( {{N}_{2}}-{{N}_{1}} \right), \\[1.1ex] 
			
			\frac{\partial {{T}_{1}}\left( x,y,t \right)}{\partial t}=g_{1}({N}_{1},{T}_{1},{Z}_{1})+{{d}_{12}}\Delta {{T}_{1}}+{{h}_{12}}\left( {{T}_{2}}-{{T}_{1}} \right), \\[1.1ex] 
			
			\frac{\partial {{Z}_{1}}\left( x,y,t \right)}{\partial t}=\varphi_{1}({N}_{1},{T}_{1},{Z}_{1})+{{d}_{13}}\Delta {{Z}_{1}}+{{d}_{31}}{{Z}_{1}}\Delta {{N}_{1}}-{{d}_{32}}{{Z}_{1}}\Delta {{T}_{1}}+{{h}_{13}}\left( {{Z}_{2}}-{{Z}_{1}} \right),\\[1.1ex]
			
			\frac{\partial {{N}_{2}}\left( x,y,t \right)}{\partial t}=f_{2}({N}_{2},{T}_{2},{Z}_{2})+{{d}_{21}}\Delta {{N}_{2}}+{{h}_{21}}\left( {{N}_{1}}-{{N}_{2}} \right), \\[1.1ex] 
			
			\frac{\partial {{T}_{2}}\left( x,y,t \right)}{\partial t}=g_{2}({N}_{2},{T}_{2},{Z}_{2})+{{d}_{22}}\Delta {{T}_{2}}+{{h}_{22}}\left( {{T}_{1}}-{{T}_{2}} \right), \\[1.1ex]
			
			\frac{\partial {{Z}_{2}}\left( x,y,t \right)}{\partial t}=\varphi_{2}({N}_{2},{T}_{2},{Z}_{2})+{{d}_{23}}\Delta {{Z}_{2}}+{{d}_{41}}{{Z}_{2}}\Delta {{N}_{2}}-{{d}_{42}}{{Z}_{2}}\Delta {{T}_{2}}+{{h}_{23}}\left( {{Z}_{1}}-{{Z}_{2}} \right), \\[1.1ex]
			
			\frac{\partial N_{i}(x,y,t)}{\partial n}=\frac{\partial T_{i}(x,y,t)}{\partial n}=\frac{\partial Z_{i}(x,y,t)}{\partial n}=0, i=1,2,(x,y)\in \partial\Omega,\\[1.1ex]
			
			N_{i}(x,y,t)>0,T_{i}(x,y,t)>0,Z_{i}(x,y,t)>0,t>0,(x,y)\in \Omega.
		\end{cases}
		\label{model}
	\end{equation}
where $\Delta=\frac{\partial ^{2}}{\partial x^{2}}+\frac{\partial ^{2}}{\partial y^{2}}$ and $\partial_n$ denotes the outward normal derivative on $\partial\Omega$. The coefficients $d_{ij}$ with $i\in\{1,2\}$ and $j\in\{1,2,3\}$ denote species-specific diffusion coefficients for $N_{i}$, $T_{i}$ and $Z_{i}$ respectively; coefficients denoted $d_{31},d_{32},d_{41},d_{42}$ represent cross-diffusion couplings between zooplankton and phytoplankton. The linear exchange terms with rates $h_{ij}$ model vertical migration between layers for each species; unless otherwise stated, we assume $h_{il}=h$ for convenience. The reaction terms employed here are Holling-type functional responses $f_{i}$, $g_{i}$ and $\varphi_{i}$ for zooplankton grazing~\cite{FeifanZhang2022,SubhenduChakraborty2012}:
\begin{equation}	
	\begin{cases}
		f_{i}({N}_{i},{T}_{i},{Z}_{i})={{r}_{i1}}{{N}_{i}}\left( 1-\frac{{{N}_{i}}+{{\alpha }_{i1}}{{T}_{i}}}{{{K}_{i1}}} \right)-\frac{{{w}_{i1}}{{N}_{i}}{{Z}_{i}}}{{{p}_{i1}}+{{N}_{i}}}, \\[1.1ex] 		
		g_{i}({N}_{i},{T}_{i},{Z}_{i})={{r}_{i2}}{{T}_{i}}\left( 1-\frac{{{T}_{i}}+{{\alpha }_{i2}}{{N}_{i}}}{{{K}_{i2}}} \right)-\frac{{{w}_{i2}}{{T}_{i}}{{Z}_{i}}}{{{p}_{i2}}+{{T}_{i}}+{{\beta }_{i}}{{N}_{i}}}, \\[1.1ex] 		
		\varphi_{i}({N}_{i},{T}_{i},{Z}_{i})=\frac{{{c}_{i1}{w}_{i1}}{{N}_{i}}{{Z}_{i}}}{{{p}_{i1}}+{{N}_{i}}}-\frac{{{c}_{i2}{w}_{i2}}{{T}_{i}}{{Z}_{i}}}{{{p}_{i2}}+{{T}_{i}}+{{\beta }_{i}}{{N}_{i}}}-{{m}_{i}}{{Z}_{i}}.\\[1.1ex]		
	\end{cases}
	\label{function}
\end{equation}
where $r_{i1}$ and $r_{i2}$ ($i=1,2$) denote the constant intrinsic growth rates of $N_{i}$ and $T_{i}$, respectively; $K_{i1}$ and $K_{i2}$ denote the carrying capacities of $N_{i}$ and $T_{i}$, respectively; $\alpha_{i1}$ and $\alpha_{i2}$ measure the competitive effect of $T_{i}$ on $N_{i}$, and $N_{i}$ on $T_{i}$, respectively; $w_{i1}$ and $w_{i2}$ represent the rates at which $N_{i}$, $T_{i}$ are consumed by $Z_{i}$, respectively; $p_{i1}$ and $p_{i2}$ denote half saturation constants for $N_{i}$ and $T_{i}$, respectively; $c_{i1}$ and $c_{i2}$ represent the conversion rates of $N_{i}$ to $Z_{i}$ and $T_{i}$ to $Z_{i}$, respectively; $\beta_{i}$ denotes the intensity of avoidance of $T_{i}$ by $Z_{i}$, in the presence of $N_{i}$; $m_{i}$ denotes the mortality rate of the zooplankton $Z_{i}$. Note that the model assumes gradient exchange between the upper and bottom layers. While this represents an idealized representation of vertical coupling that may not capture all complexities of real stratified ecosystems, it provides a minimal framework for isolating the fundamental role of diffusive coupling in shaping spatial dynamics and noise robustness.
\begin{figure*}[ht!]
	\centering
	\includegraphics[width= 17cm]{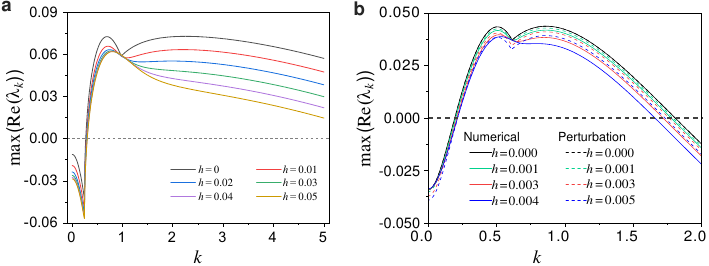}
	\caption{\label{sesanquxian1-4} 
		\textbf{Dispersion relations and validation of weak-coupling approximation.}
		\textbf{(a)} Maximum real part of the eigenvalues, $\max \mathrm{Re}(\lambda_k)$, versus wavenumber $k$ for representative interlayer coupling strengths $h$. Increasing $h$ shifts and merges the two independent Turing peaks, indicating synchronization of pattern-forming modes across layers. 
		\textbf{(b)} Comparison between direct numerical solutions (solid lines) and first-order perturbation predictions (dashed lines) for several small coupling strengths ($h\le0.005$). The black curve ($h=0$) represents the non-coupled case. The excellent agreement validates the perturbative analysis in the weak-coupling limit (see Eq. S8). Curves correspond to parameter sets in Table~S1 of the Supplementary Materials (SM).}
\end{figure*}
\subsection{Independent pattern formation in the non-coupled system}	
We first analyze the non-coupled case~(see Fig.~\ref{sesanquxian1-4} for $h=0$), which serves as a reference regime for understanding the intrinsic pattern-forming tendencies of each layer. Uniform steady states are obtained by setting all temporal and spatial derivatives to zero, yielding a coexistence equilibrium $E^{*}$=($N_1^*$, $T_1^*$, $Z_1^*$, $N_2^*$, $T_2^*$, $Z_2^*$) in which all species persist in both layers. Linear stability analysis around $E^{*}$ leads to a dispersion relation $\max(\text{Re}(\lambda_k))$ across a range of wavenumber $k$. For each $k$, six eigenvalues arise from the linearized system, reflecting the three-species dynamics in two non-coupled layers~(see Supplementary Materials (SM)~I for details). According to Turing instability theory~\cite{AMTuring1952,Murray2003,XiaoChongZhang2012,LinXue2012}, spatial patterning occurs when $\max(\text{Re}(\lambda_{k})) > 0$ for some $k > 0$. 

\begin{figure*}[ht!]
	\centering
	\includegraphics[width=14cm]{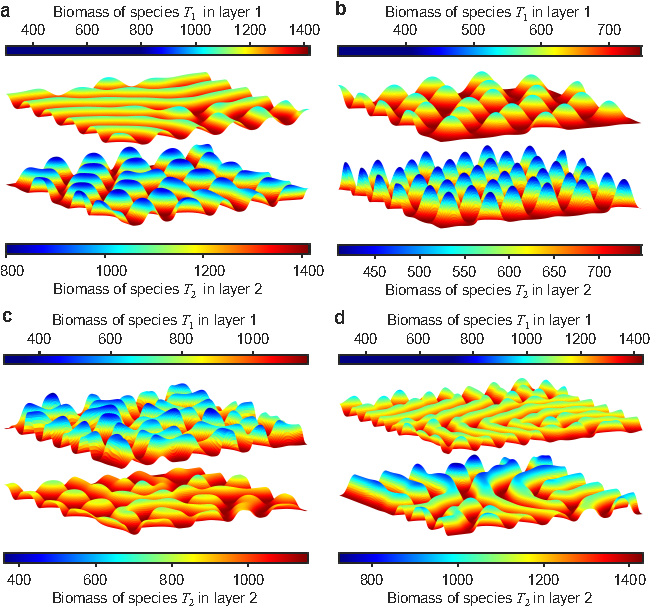}
	\caption{\label{3DpatternNonSych} 
		Self-organized pattern dynamics in the non-coupled aquatic system (without vertical migration, $h = 0$). (a-d) Spatial patterns in the surface (\textit{layer} 1) and bottom (\textit{layer} 2) layers exhibit distinct structural forms. Under different parameter sets but identical initial conditions, the system exhibits a variety of coexisting spatial configurations, each with distinct pattern types in the two \textit{layers}: (a) stripes in \textit{layer} 1 and spots in \textit{layer} 2, (b) spots in both \textit{layers}, (c) mixed patterns in both \textit{layers}, and (d) stripes in both \textit{layers}. The dispersion relation corresponding to the case in (a) is shown in Fig.~\ref{sesanquxian1-4}; panels (b-d) exhibit analogous dispersion behaviour (not shown). The spatial pattern structures of non-toxic phytoplankton ($N_i$) and zooplankton ($Z_i$) closely resemble those of toxic phytoplankton ($T_i$) (see Fig.~S1). The parameter values are summarized in Table~S1 of the SM.} 	
\end{figure*}

In the absence of coupling, the dispersion relation exhibits two distinct peaks (see Fig.~\ref{sesanquxian1-4}a, $h=0$), each corresponding to a critical Turing mode localized in each layer. This double-peak structure indicates that both layers independently satisfy the conditions for diffusion-driven instability, but with different preferred spatial scales. Numerical simulations confirm this prediction: the upper (layer 1) and lower (layer 2) layers develop self-organized patterns with distinct spatial structures, including spots, stripes, and mixed forms (Figs.~\ref{3DpatternNonSych},~S1). These differences arise from layer-specific parameter values, such as diffusion coefficients and interaction strengths, which selectively amplify different instability modes. The non-coupled system therefore supports independent pattern formation in each layer, without any tendency toward spatial synchronization.

To quantitatively verify the perturbative treatment in the weak-coupling regime, we compare the numerical solution of the full linearized system with the first-order approximation derived from Rayleigh-Schrödinger perturbation theory (see SM Sec. II). As shown in Fig.~\ref{sesanquxian1-4}b, the perturbative prediction (dashed lines) closely matches the numerical result (solid lines) for all tested coupling strengths in the range $h\le0.005$, confirming the validity of the analytical expansion. This agreement substantiates the interpretation that the peak merging observed in Fig.~\ref{sesanquxian1-4}a originates from the coupling-induced shift of the individual layer modes.
\subsection{Coupling-induced synchronization of spatial patterns}
\begin{figure*}[ht!]
	\centering
	\includegraphics[width=14cm]{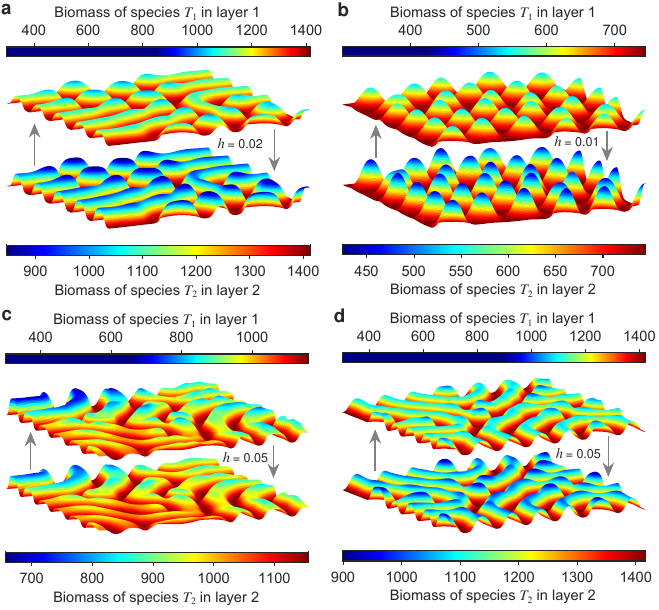}
	\caption{\label{3DpatternSych} 
		Self-organized synchronization pattern dynamics in the coupled aquatic system with vertical migration. (a-d) With the introduction of coupling, the initially independent spatial patterns from Fig.~\ref{3DpatternNonSych} (e.g., spots, stripes, mixtures) evolve into synchronized states across layers. The merging of the two dispersion-relation peaks at $h = 0$ into a single peak at $h = 0.02$ (see Fig.~\ref{sesanquxian1-4}) corresponds to this spatiotemporal synchronization shown in Fig.~\ref{3DpatternSych}a, and Fig.~\ref{3DpatternSych}b-d exhibit analogous synchronous behavior. The spatial pattern structures of non-toxic phytoplankton ($N_i$) and zooplankton ($Z_i$) closely resemble those of toxic phytoplankton ($T_i$) (see Fig.~S2).
		The parameter values are summarized in Table~S1 of the SM.} 	
\end{figure*}

Introducing inter-layer coupling fundamentally alters the pattern-forming dynamics. As the coupling strength $h$ increases from zero, the two initially separated peaks in the dispersion relation gradually approach each other and eventually merge into a single dominant peak~(see Fig.~\ref{sesanquxian1-4}a, $h>0$). This transition indicates that the previously independent Turing modes become dynamically linked through vertical migration. The simulations confirm this dynamical shift, as the initially asynchronous layers evolve toward a state of synchronized spatial patterns~(see Figs.~\ref{3DpatternSych},~S2). Patterns that were distinct in the non-coupled regime evolve into spatially coherent structures sharing the same dominant wavelength and morphology. This synchronization reflects the locking of pattern-forming modes between layers, driven by sufficient inter-layer exchange.

To test the generality of the synchronization result, we performed a systematic scan in which the cross-diffusion coefficient $d_{32}$, which couples zooplankton to gradients of toxic phytoplankton, was varied while all other parameters were held fixed (Fig.~S3). At every $d_{32}$ the layer difference decreases systematically with the coupling strength, the steepest decrease occurring around the synchronization threshold, and at $h=0.1$ the error is one to two orders of magnitude smaller than in the uncoupled case. Because this decrease occurs at essentially the same coupling strength for every $d_{32}$, the threshold does not require a fine tuning of the cross-diffusion coefficient. The scan moreover uses a fifth parameter set, listed in the last row of Table~S1, which differs from the four sets of the main text in its growth rates, grazing rates and cross-diffusion coefficients, so the synchronization reported here is not tied to a particular parameter combination. The four parameter sets presented in the main text were selected to illustrate the different pattern morphologies (spots, stripes, and mixed patterns) that arise in different parameter regimes, rather than to sample the parameter space.

\begin{figure*}[ht!]
	\centering
	\includegraphics[width=14cm]{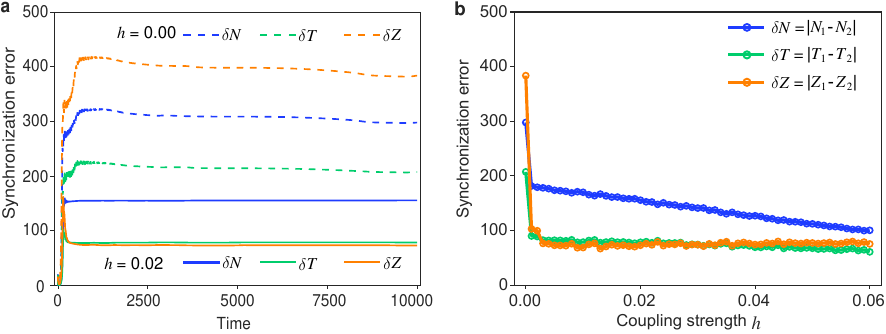}
	\caption{\label{sync_error_h} 
		Synchronization error analysis for the coupled aquatic system with vertical migration corresponding to Figs.~\ref{3DpatternNonSych}a,~\ref{3DpatternSych}a. (a) Temporal evolution of the synchronization error for non-coupled ($h = 0$) and coupled ($h = 0.02$) conditions. The error increases and plateaus without coupling, whereas it decreases and plateaus with coupling.  The steady-state synchronization error in the non-coupled system is 3 to 5 times larger than that in the coupled system. (b) Steady-state synchronization error as a function of coupling strength $h$. The error decreases monotonically with increasing $h$ and eventually stabilizes.
		The parameter values are summarized in Table~S1 of the SM.} 	
\end{figure*}

\begin{figure*}[ht!]
	\centering
	\includegraphics[width=16cm]{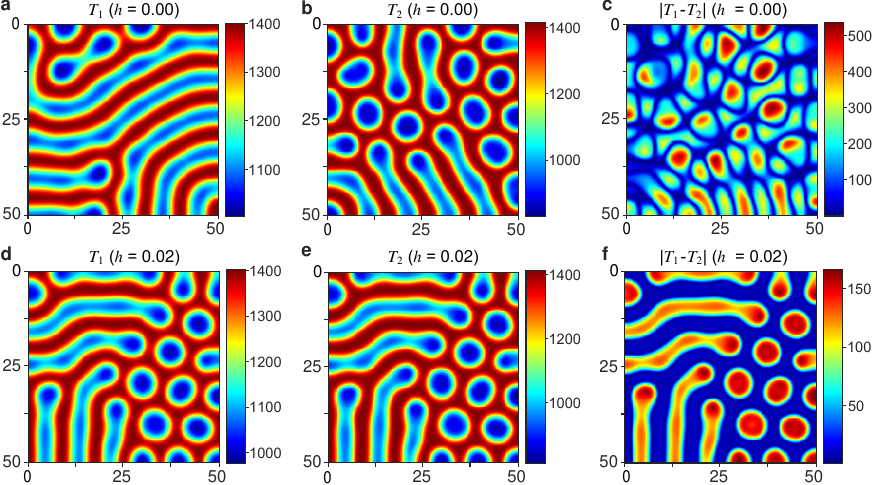}
	\caption{\label{sync_error_Fig} 
		Visual comparison of self-organized patterns and their synchronization errors for non-coupled (Fig.~\ref{3DpatternNonSych}a) and coupled (Fig.~\ref{3DpatternSych}a) systems. (a-c) For the non-coupled system ($h = 0$): (a-b) self-organized patterns, (c) pattern of the synchronization error, which exhibits a structure distinctly different from (a-b).  (d-f) For the coupled system ($h = 0.02$): (d-e) synchronized patterns, (f) pattern of the synchronization error, which maintains structural similarity to (d-e). The spatial pattern structures of non-toxic phytoplankton ($N_i$) and zooplankton ($Z_i$) closely resemble those of toxic phytoplankton ($T_i$) (see Fig.~S4).
		The parameter values are summarized in Table~S1 of the SM.} 	
\end{figure*}

The synchronization transition is quantified by defining the synchronization errors as $\delta N(x,y,t)= \left|N_{1}(x,y,t)-N_{2}(x,y,t) \right| $, $\delta T(x,y,t)= \left|T_{1}(x,y,t)-T_{2}(x,y,t) \right| $, and $\delta Z(x,y,t)= \left|Z_{1}(x,y,t)-Z_{2}(x,y,t) \right| $, where the differences are evaluated pointwise at each spatial location $(x,y)$ within the domain. In the non-coupled regime, these errors remain large and fluctuate around elevated steady values~(see Fig.~\ref{sync_error_h}a). When coupling is introduced, the errors decrease over time and converge to substantially smaller steady-state levels. The steady-state synchronization error in the non-coupled system is approximately three to five times larger than that in the coupled system. This indicates that non-synchronized patterns maintain their asynchronous characteristics even after processing synchronization errors~(Figs.~\ref{sync_error_Fig}a-c,~S4), while synchronized patterns remain coherent under the same treatment (Figs.~\ref{sync_error_Fig}d-f,~S4).

The dependence of synchronization error on coupling strength reveals a critical coupling threshold $h_c$~(see Fig.~\ref{sync_error_h}b). For $h < h_c$, synchronization remains incomplete, whereas for $h > h_c$ the system consistently converges to a synchronized pattern state. This behavior demonstrates that inter-layer coupling can overcome intrinsic heterogeneities and enforce coherent spatial organization across stratified layers.

\subsection{Effects of stochasticity on pattern formation}	

\begin{figure*}[ht!]
	\centering
	\includegraphics[width=16cm]{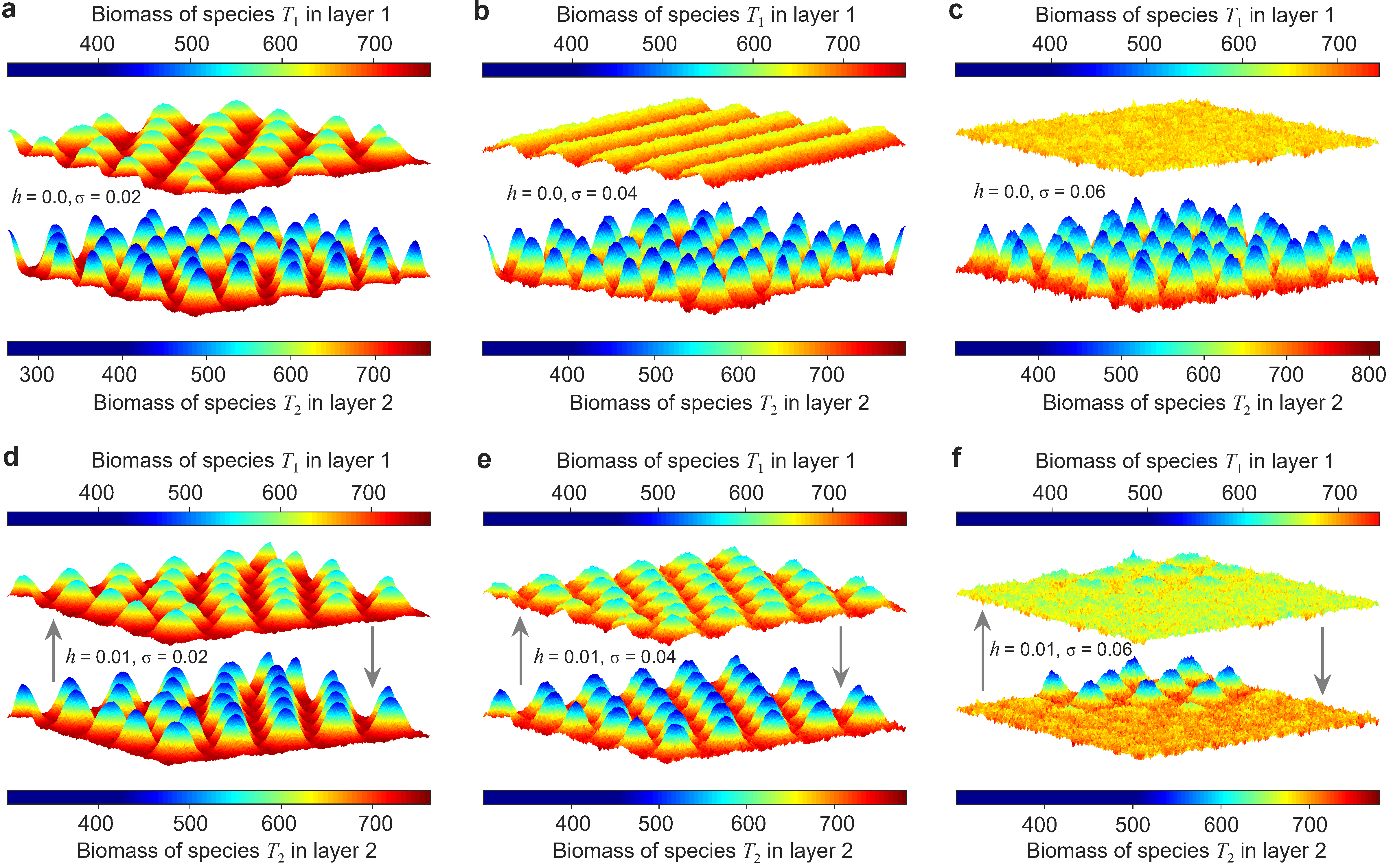}
	\caption{\label{Figdot} 
		Effect of stochasticity on spotted pattern structures. (a-c) Without coupling, increasing noise intensity affects the upper layer more than the lower layer: the upper-layer pattern transitions from spots to stripes and eventually vanishes, while the lower layer maintains its spotted structure. (d-f) With coupling, the system exhibits robustness to stochastic fluctuations, preserving the spotted structure in both layers until the noise intensity exceeds a critical threshold, beyond which the spatial patterns collapse. The spatial pattern structures of non-toxic phytoplankton ($N_i$) and zooplankton ($Z_i$) closely resemble those of toxic phytoplankton ($T_i$) (see Fig.~S5). The parameter values are summarized in Table~S1 of the SM.} 	
\end{figure*}	
Natural aquatic environments are subject to persistent environmental fluctuations, motivating an examination of stochastic effects on pattern dynamics. We therefore introduce multiplicative Gaussian white noise into the model and analyze its impact on spatial structures in both non-coupled and coupled regimes.

In the non-coupled system, stochastic perturbations induce markedly different responses across layers. As noise intensity increases, the upper layer undergoes a progressive transition from spot-like patterns to stripe-like structures and eventually to pattern collapse, whereas the lower layer retains its spotted configuration over the same range of noise intensities~(see Figs.~\ref{Figdot}a-c,~S5). This contrast indicates a layer-dependent sensitivity to stochastic forcing.

When coupling is present, synchronized patterns persist under moderate noise levels and remain qualitatively similar to those observed in deterministic simulations. At sufficiently high noise intensities $\sigma = 0.06$, synchronization degrades and spatial structures become increasingly disordered, with the upper layer again exhibiting more pronounced disruption~(see Figs.~\ref{Figdot}d-f,~S5). This asymmetric response indicates that sufficiently strong stochasticity can disrupt the delicate balance maintaining self-organized dynamics, ultimately destroying the coherent spatial structures generated through Turing mechanisms. Our results show that strong stochasticity can destabilize self-organized patterns, even in the presence of coupling, suggesting the crucial role of environmental variability in shaping spatial ecology.

The noise robustness of the synchronized state can be given an analytical basis through the linear response of the layer-difference mode (SM Sec.~VI). In the symmetric-antisymmetric basis the coupling leaves the wavenumber dependence of the linearized operator unchanged and shifts the antisymmetric block by $-2h$ [Eq.~(S10)], so that the component measured by the synchronization errors is damped at a rate that grows with the coupling strength. When the inter-layer mismatch $\delta_0$ between the two layer modes is smaller than $2h$, which is precisely the regime in which the two dispersion peaks have merged, the antisymmetric component becomes a damped Ornstein-Uhlenbeck process, and its noise-driven part scales as $\sigma(2h-\delta_0)^{-1/2}$. The damping therefore acts directly on the quantity that the synchronization error measures, which accounts for the monotonic decrease of the steady-state error with $h$ in Fig.~\ref{sync_error_h}b. Varying the noise intensity at fixed coupling is consistent with this picture. Fig.~S6 shows the three synchronization errors as functions of $\sigma$ for nine coupling strengths between $h=0$ and $h=0.1$: for $\sigma\le0.06$ the error decreases with $h$ for all three species, while at larger $\sigma$ the patterns are strongly degraded and the errors no longer vary monotonically. Among the three species, $\langle|Z_{1}-Z_{2}|\rangle$ is the one that grows most steeply with $\sigma$. The estimate describes only this noise-driven part: at small coupling the measured error is dominated by the deterministic mismatch between the two layers, so it does not by itself determine the noise intensity at which the pattern collapses, and a closed-form expression for that threshold would require the nonlinear amplitude equations of the coupled system.

\begin{figure*}[ht!]
	\centering
	\includegraphics[width=16cm]{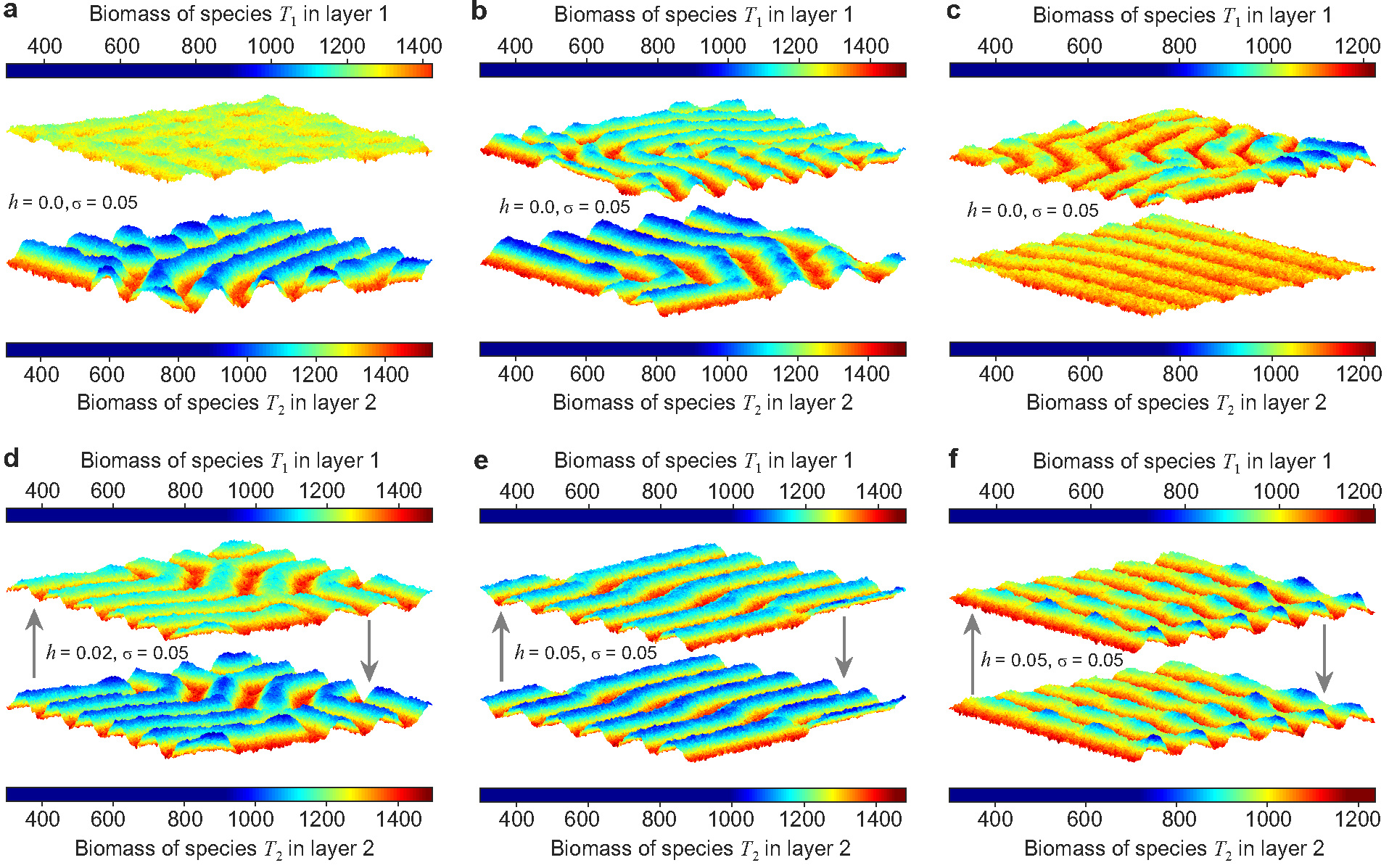}
	\caption{\label{Fignoise} 
		Robust synchronization patterns by coupling in a stochastic environment. (a-c) Without coupling, the two layers maintain distinct and independent spatial patterns in the presence of noise. (d-f) With coupling introduced, the initially different patterns achieve a robust synchronized state, demonstrating that coupling can overcome both initial heterogeneity and stochastic perturbations to establish synchronous dynamics. The spatial pattern structures of non-toxic phytoplankton ($N_i$) and zooplankton ($Z_i$) closely resemble those of toxic phytoplankton ($T_i$) (see Fig.~S7). The parameter values are summarized in Table~S1 of the SM.} 	
\end{figure*}

\subsection{Robustness of synchronized patterns under stochastic perturbations}	
Despite the disruptive effects of noise, the coupled system exhibits a pronounced robustness compared to the non-coupled case. While stochasticity leads to substantial asymmetry and degradation of patterns in non-coupled layers, coupling maintains coherent and synchronized spatial structures across layers over a broad range of noise intensities~(see Figs.~\ref{Fignoise},~S7).

The impact of noise is also species-dependent. At a noise intensity of $\sigma=0.05$, the spatial distributions of both phytoplankton types maintain relatively stable configurations, whereas zooplankton displays pronounced local fluctuations, particularly in the upper layer. Simulations indicate that non-toxic phytoplankton is least affected by stochastic perturbations, toxic phytoplankton shows intermediate sensitivity, and zooplankton exhibits the strongest response, particularly in the upper layer. This hierarchy suggests that higher trophic levels are more susceptible to environmental fluctuations. Importantly, coupling mitigates these layer-specific effects by redistributing disturbances between layers. Zooplankton patterns that fragment under non-coupled conditions recover coherent spatial organization when coupling is introduced. Overall, inter-layer coupling enhances both synchronization and resilience, stabilizing self-organized patterns against stochastic perturbations.

\section{Discussion}	
The persistence of coherent spatial patterns in plankton ecosystems, despite pervasive environmental variability, represents a fundamental challenge for understanding self-organization in aquatic systems~\cite{SubhenduChakraborty2012,FeifanZhang2022,ToushengHuangCNSNS,ToushengHuang2023,CaroLFolt1999,LevinSA1976,AlexanderMedvinsky2002,JenniferPrairie2012,Edward R.Abraham1998,WilliamMDurham2013,APMartin2003}. Although a wide range of biological and physical mechanisms have been proposed to explain plankton patchiness~\cite{ToushengHuangCNSNS,ToushengHuang2023,BingzhangChen2020}, far less attention has been paid to how spatially synchronized patterns remain stable once coupling between domains is established, particularly under stochastic conditions. In this study, we address this issue by demonstrating that gradient coupling between vertically structured plankton layers provides a minimal yet effective mechanism for both synchronization and noise stability. 

A central result of our work is the identification of a sharp, coupling-induced transition from layer-specific pattern formation to fully synchronized spatial structures. In the absence of coupling, each layer independently develops Turing-type patterns that are highly susceptible to stochastic perturbations. Once coupling strength exceeds a critical threshold, however, these independent patterns collapse into a unified spatial configuration that remains coherent even under substantial environmental noise. This finding establishes coupling strength as a key control parameter governing the stability and coherence of pattern-forming plankton systems, rather than merely a secondary modifier of pre-existing patterns.

Beyond synchronization, our results reveal that diffusive coupling fundamentally reshapes how environmental noise propagates through the ecosystem. In particular, we uncover a systematic trophic hierarchy in noise sensitivity, with zooplankton populations exhibiting markedly greater vulnerability to stochastic perturbations than phytoplankton. The mechanistic origin of this hierarchy can be understood from the structure of the interaction terms. Zooplankton populations depend on two prey types (toxic and non-toxic phytoplankton) for growth, making them exposed to fluctuations in both prey populations, whereas phytoplankton dynamics are primarily governed by their own growth and competitive interactions. This asymmetry, combined with the fact that zooplankton have higher effective diffusion coefficients (including cross-diffusion terms) that amplify spatial heterogeneity, explains why higher trophic levels exhibit greater vulnerability to stochastic perturbations. A full analytical characterization of this hierarchy would require a stochastic stability analysis of the coupled system, which represents an important direction for future theoretical work. Such differential sensitivity may contribute to trophic imbalances and may offer mechanistic insight into why higher trophic levels often exhibit earlier or more pronounced responses to environmental variability.

Importantly, the mechanism identified here differs qualitatively from synchronization driven by active biological processes, such as zooplankton vertical migration or adaptive foraging behavior~\cite{JenniferPrairie2012,CaroLFolt1999,BingzhangChen2020,KendraLDaly1993,AndrewYuMorozov2007}. Whereas these processes rely on species-specific traits and behavioral responses, diffusive coupling is a passive physical mechanism arising from the exchange of individuals or biomass between layers. Our results therefore demonstrate that robust synchronization can emerge even in the absence of complex behavioral regulation, provided that sufficient coupling exists between spatial domains. This distinction clarifies the respective roles of physical coupling and biological adaptation and suggests that some large-scale spatial organization in plankton communities may be explained without invoking finely tuned behavioral strategies.

The ecological significance of synchronization becomes apparent when considering how real stratified ecosystems function. The convergence of upper and lower layers toward coherent spatial patterns does not negate the importance of stratification; rather, it reveals that vertical migration, as exemplified by organisms such as euphausiids (\textit{Euphausia mucronata}) and semi-pelagic squat lobsters (\textit{Pleuroncodes monodon}) in oxygen minimum zones, creates essential functional connectivity across physically distinct habitats \cite{RainerKiko2019}. This biological coupling transforms independent dynamical systems into an integrated ecological unit, where synchronized patterns reflect coordinated responses to environmental variability across the water column. Here, stratification provides the fundamental template for niche differentiation, while synchronization ensures system-level coherence. Importantly, synchronization confers measurable benefits that directly address ecosystem stability: global-scale analyses of plankton community interactomes reveal that coordinated dynamics enhance robustness to environmental perturbations, with more tightly connected communities exhibiting lower vulnerability to changing conditions \cite{Chaffron2021}. Our finding that coupled systems exhibit enhanced robustness to stochastic perturbations aligns precisely with this empirical evidence, suggesting that coherent spatial organization actively contributes to ecosystem resilience. 

Furthermore, the fact that layers initially support different types of pattern demonstrates that each layer retains its intrinsic self-organizing capacity, reflecting the functional diversity observed in mixotrophic plankton communities, where flexible foraging strategies allow organisms to occupy distinct ecological niches \cite{Ward2019}. Their eventual synchronization under coupling reveals how migratory integration weaves these local dynamics into a coherent whole without erasing underlying heterogeneity. This interplay between local heterogeneity and regional coherence has direct implications for understanding ecosystem functioning: the maintenance of distinct layer-specific dynamics prior to synchronization supports the stoichiometric diversity observed across trophic levels, which underpins energy transfer efficiency and population persistence under environmental stress \cite{Feng2023}. Thus, stratification provides the necessary spatial structure for functional specialization, while biologically mediated synchronization ensures that these specialized components operate in concert, buffering the system against localized disturbances, and promoting long-term stability.

From a theoretical perspective, our findings place the classical Turing instability in a broader ecological context. Our work extends the established framework of Turing instabilities in two-layer networks~\cite{Asllani2014,CalderonBarreto2026} by demonstrating that inter-layer diffusion not only enables pattern synchronization but also fundamentally alters the stability of these patterns under stochastic perturbations. While prior work has characterized the linear stability conditions for pattern formation in coupled networks, our results reveal how coupling strength, noise intensity, and trophic structure jointly govern the robustness of self-organized patterns in fluctuating environments. For the layer difference this interplay can be written explicitly: the linear-response estimate of SM Sec.~VI gives a stationary synchronization error proportional to $\sigma(2h-\delta_0)^{-1/2}$, so that the layer-difference mode is stable only above $h_c$ and the noise-driven part of the synchronization error is suppressed as $h$ grows. Rather than acting as a global system-wide property, pattern-forming instabilities operate locally within layers, while inter-layer coupling determines whether these patterns remain fragmented or become synchronized across space. Environmental noise further modulates this balance, selectively destabilizing non-coupled configurations while leaving coupled systems comparatively robust. This interplay between local instability, coupling, and stochasticity offers a unifying framework for understanding why coherent plankton patterns are frequently observed in noisy aquatic environments. 

The framework presented here is intentionally minimal and does not explicitly incorporate adaptive behaviors, species diversity beyond a simple trophic structure, or complex hydrodynamic forcing. In natural ecosystems, these factors are likely to interact with diffusive coupling in nontrivial ways. Future work should therefore explore how passive coupling interacts with active migration, adaptive foraging, and higher trophic complexity, as well as how the noise robustness identified here extends to more diverse food webs~\cite{JuKangeLife,JuKangCSF,JuKangCSF2026}. Nonetheless, by isolating diffusive coupling as a core organizing mechanism, our results provide a parsimonious baseline for interpreting the emergence and persistence of coherent spatial patterns in fluctuating aquatic ecosystems.

\section{Methods}
We investigated the spatiotemporal dynamics of a spatially explicit, two-layer aquatic ecosystem model describing interactions among non-toxic phytoplankton ($N_i$), toxic phytoplankton ($T_i$), and zooplankton ($Z_i$) distributed across two vertically stratified water layers ($i = 1,2$). The system is formulated as a set of coupled reaction-diffusion equations incorporating species-specific diffusion, biologically motivated cross-diffusion, and inter-layer exchange representing vertical migration.

\subsection{Numerical simulations}
\textbf{Computational domain and discretization}.
All simulations were conducted on a two-dimensional square domain of size $50 \times 50$, representing a horizontally extended aquatic habitat. The domain was discretized using a uniform Cartesian grid with spatial steps $\delta x = \delta y = 0.25$, yielding a total of $201 \times 201$ grid points. This spatial resolution was sufficient to resolve the characteristic wavelengths of the emerging Turing patterns observed in the system.\\
\textbf{Time integration}.
Temporal evolution was computed using an explicit forward Euler scheme with a fixed time step $\delta t = 0.01$. This choice ensures numerical stability and adequate temporal resolution for the reaction-diffusion dynamics under the parameter regimes considered. Each simulation was run for 10,000 time steps, allowing transient behavior to decay and long-term spatiotemporal patterns to fully develop.\\
\textbf{Boundary conditions}.
Zero-flux (von Neumann) boundary conditions were imposed on all state variables at the domain boundaries, ensuring that no population flux occurs across the system edges. These conditions correspond to an isolated ecosystem and conserve total biomass within the computational domain.\\
\textbf{Stochastic framework}.
To account for environmental variability, stochastic perturbations were incorporated in the form of multiplicative Gaussian white noise applied to all population variables. At each time step, noise terms with intensity $\sigma$ were added and scaled by $\sqrt{\delta t}$ to ensure appropriate temporal scaling. The stochastic simulations were implemented using a Monte Carlo approach. To maintain biological realism and numerical stability, population densities were constrained to remain non-negative throughout the simulations.\\
\textbf{Initial conditions}.
Simulations were initialized with spatially heterogeneous population distributions to reflect natural patchiness in aquatic ecosystems and to allow spontaneous symmetry breaking. Initial biomasses were set as spatially uniform backgrounds with superimposed random perturbations: $N_{i}(0)=500 \pm 100$, $T_{i}(0)=300 \pm 50$, and $Z_{i}(0)=100 \pm 20$ across the domain. These initial conditions do not impose any predefined spatial structure and serve solely to trigger pattern-forming instabilities.

\textbf{Numerical validation}.
	To ensure the reliability of our numerical results, we performed several validation tests. First, we verified that the numerical solutions converged with spatial and temporal resolution by repeating selected simulations with $\delta x = \delta y = 0.125$ (twice the resolution) and $\delta t = 0.005$ (half the time step). The resulting pattern morphologies and synchronization errors differed by less than 2\%, confirming that the chosen resolution ($201 \times 201$ grid points, $\delta t = 0.01$) is sufficient. Second, we compared the numerically observed pattern wavelengths with the theoretical predictions from the dispersion relations (Fig. 1) and found excellent agreement. Third, the stochastic simulations were performed for a single realization of the noise for each parameter set, starting from the same type of random initial conditions as the deterministic runs.

\subsection{Model processes}
\textbf{Biological Interactions}.
The model incorporates several key ecological processes: (1) logistic population growth with inter-specific competition between phytoplankton types, (2) type-II functional response for zooplankton grazing, (3) biomass conversion with distinct efficiencies for toxic and non-toxic prey, (4) density-dependent mortality, and (5) vertical migration between water layers.\\
\textbf{Spatial processes}.
Population dispersal within each layer is described by standard diffusion terms with species-specific diffusion coefficients. In addition, cross-diffusion terms are included to model directed zooplankton movement in response to phytoplankton gradients, capturing avoidance of toxic phytoplankton and attraction toward non-toxic prey. These terms provide a mechanistic representation of behaviorally mediated movement and play a critical role in the emergence of spatial patterns.

\section*{Acknowledgments}
We thank Yuanzhi Li for helpful discussions. This study was funded by the National Natural Science Foundation of China (32525006, 32330064), and the Fundamental and Interdisciplinary Disciplines Breakthrough Plan of the Ministry of Education of China (JYB2025XDXM902). 

\section*{Conflict of interest}
The authors declare that they have no known competing financial interests or personal relationships that could have appeared to influence the work reported in this paper.

\section*{Author contributions}
J.K. and C.C conceived the project and planned the study. All authors supervised the study, developed the model, analyzed the data, carried out the theoretical analysis and numerical simulations, analyzed the simulation results, and wrote the paper. 

\section*{Data and code availability}
No data was used for the research described in this paper.

	
	\newpage
	\textbf{\LARGE{Supplementary Materials}}
	
	\renewcommand \thesection {\Roman{section}}
	\renewcommand \thesubsection {\Alph{subsection}}
	\renewcommand \thesubsubsection {\arabic{subsubsection}}
	
	\makeatletter
	\renewcommand{\thefigure}{S\@arabic\c@figure}
	\renewcommand{\theequation}{S\@arabic\c@equation}
	\makeatother
	
	\setcounter{section}{0}    
	\setcounter{figure}{0}    
	\setcounter{equation}{0}
	\renewcommand\theequation{S\arabic{equation}}		
\section {Turing bifurcation analysis of the coupled system }\label{Turingbifurcationanalysis}	
In this work, we investigate the reaction-diffusion systems in coupled spatiotemporal aquatic ecosystems characterized by vertical migration as follows: 
\begin{equation}
	\begin{scriptsize}	
		\begin{cases}
			\frac{\partial {{N}_{1}}\left( x,y,t \right)}{\partial t}={{r}_{11}}{{N}_{1}}\left( 1-\frac{{{N}_{1}}+{{\alpha }_{11}}{{T}_{1}}}{{{K}_{11}}} \right)-\frac{{{w}_{11}}{{N}_{1}}{{Z}_{1}}}{{{p}_{11}}+{{N}_{1}}}+{{d}_{11}}\Delta {{N}_{1}}+{{h}_{11}}\left( {{N}_{2}}-{{N}_{1}} \right),(x,y)\in \Omega, t>0, \\[1.1ex] 
			
			\frac{\partial {{T}_{1}}\left( x,y,t \right)}{\partial t}={{r}_{12}}{{T}_{1}}\left( 1-\frac{{{T}_{1}}+{{\alpha }_{12}}{{N}_{1}}}{{{K}_{12}}} \right)-\frac{{{w}_{12}}{{T}_{1}}{{Z}_{1}}}{{{p}_{12}}+{{T}_{1}}+{{\beta }_{1}}{{N}_{1}}}+{{d}_{12}}\Delta {{T}_{1}}+{{h}_{12}}\left( {{T}_{2}}-{{T}_{1}} \right),(x,y)\in \Omega, t>0, \\[1.1ex] 
			
			\frac{\partial {{Z}_{1}}\left( x,y,t \right)}{\partial t}=\frac{{{c}_{11}{w}_{11}}{{N}_{1}}{{Z}_{1}}}{{{p}_{11}}+{{N}_{1}}}-\frac{{{c}_{12}{w}_{12}}{{T}_{1}}{{Z}_{1}}}{{{p}_{12}}+{{T}_{1}}+{{\beta }_{1}}{{N}_{1}}}-{{m}_{1}}{{Z}_{1}}+{{d}_{13}}\Delta {{Z}_{1}}+{{d}_{31}}{{Z}_{1}}\Delta {{N}_{1}}-{{d}_{32}}{{Z}_{1}}\Delta {{T}_{1}}+{{h}_{13}}\left( {{Z}_{2}}-{{Z}_{1}} \right),(x,y) \in \Omega, t>0,\\[1.1ex]
			
			\frac{\partial {{N}_{2}}\left( x,y,t \right)}{\partial t}={{r}_{21}}{{N}_{2}}\left( 1-\frac{{{N}_{2}}+{{\alpha }_{21}}{{T}_{2}}}{{{K}_{21}}} \right)-\frac{{{w}_{21}}{{N}_{2}}{{Z}_{2}}}{{{p}_{21}}+{{N}_{2}}}+{{d}_{21}}\Delta {{N}_{2}}+{{h}_{21}}\left( {{N}_{1}}-{{N}_{2}} \right),(x,y)\in \Omega, t>0, \\[1.1ex] 
			
			\frac{\partial {{T}_{2}}\left( x,y,t \right)}{\partial t}={{r}_{22}}{{T}_{2}}\left( 1-\frac{{{T}_{2}}+{{\alpha }_{22}}{{N}_{2}}}{{{K}_{22}}} \right)-\frac{{{w}_{22}}{{T}_{2}}{{Z}_{2}}}{{{p}_{22}}+{{T}_{2}}+{{\beta }_{2}}{{N}_{2}}}+{{d}_{22}}\Delta {{T}_{2}}+{{h}_{22}}\left( {{T}_{1}}-{{T}_{2}} \right),(x,y)\in \Omega, t>0, \\[1.1ex]
			
			\frac{\partial {{Z}_{2}}\left( x,y,t \right)}{\partial t}=\frac{{{c}_{21}{w}_{21}}{{N}_{2}}{{Z}_{2}}}{{{p}_{21}}+{{N}_{2}}}-\frac{{{c}_{22}{w}_{22}}{{T}_{2}}{{Z}_{2}}}{{{p}_{22}}+{{T}_{2}}+{{\beta }_{2}}{{N}_{2}}}-{{m}_{2}}{{Z}_{2}}+{{d}_{23}}\Delta {{Z}_{2}}+{{d}_{41}}{{Z}_{2}}\Delta {{N}_{2}}-{{d}_{42}}{{Z}_{2}}\Delta {{T}_{2}}+{{h}_{23}}\left( {{Z}_{1}}-{{Z}_{2}} \right),(x,y)\in \Omega, t>0, \\[1.1ex]
			
			\frac{\partial N_{i}(x,y,t)}{\partial n}=\frac{\partial T_{i}(x,y,t)}{\partial n}=\frac{\partial Z_{i}(x,y,t)}{\partial n}=0, i=1,2,(x,y)\in \partial\Omega, t>0,\\[1.1ex]
			
			N_{i}(x,y,t)>0,T_{i}(x,y,t)>0,Z_{i}(x,y,t)>0,(x,y)\in \Omega.
		\end{cases}
		\label{Smodel}
	\end{scriptsize}
\end{equation}
Here, from the ecological perspective, our focus is on the coexistence state $E^{*}$=($N_1^*$, $T_1^*$, $Z_1^*$, $N_2^*$, $T_2^*$, $Z_2^*$) of the system~(\ref{Smodel}) in both layers. As established, vertical environmental heterogeneity is modeled through inter-layer parameter variations. Theoretically, the impact of varying any parameter subset on the pattern formation in system~(\ref{Smodel}) can be investigated. 

In addition, the non-spatial coexistence steady state ($N_1^*$, $T_1^*$, $Z_1^*$) of a non-coupled single system is well-established in previous research~\cite{SubhenduChakraborty2012,FeifanZhang2022}.  
Hence, we are interested in studying the stability of the coexisting equilibrium $E^{*}$ within coupled system~(\ref{Smodel}). Hence, the Jacobian matrix of the non-spatial form of system~(\ref{Smodel})
associated with $E^{*}$=($N_1^*$, $T_1^*$, $Z_1^*$, $N_2^*$, $T_2^*$, $Z_2^*$) can be obtained as follows
\begin{equation}
	J=\begin{pmatrix}
		a_{11} & a_{12} & a_{13} & a_{14} & 0      & 0 \\
		a_{21} & a_{22} & a_{23} & 0      & a_{25} & 0 \\
		a_{31} & a_{32} & a_{33} & 0      & 0      & a_{36}\\
		a_{41} & 0      & 0      & a_{44} & a_{45}      & a_{46} \\
		0      & a_{52} & 0      & a_{54} & a_{55}      & a_{56} \\
		0      & 0      & a_{63} & a_{64} & a_{65}      & a_{66} 		
	\end{pmatrix},
	\label{Jacobian}
\end{equation}
where $a_{11} = -h_{11} - \frac{p_{11} w_{11} Z^{*}_{1}}{(N^{*}_{1} + p_{11})^2} + \frac{r_{11} (K_{11} - 2 N^{*}_{1} - T^{*}_{1} \alpha_{11})}{K_{11}}$,
$a_{12} = -\frac{N^{*}_{1} r_{11} \alpha_{11}}{K_{11}}$,
$a_{13} = -\frac{N^{*}_{1} w_{11}}{N^{*}_{1} + p_{11}}$,
$a_{14} = h_{11}$,
$a_{21} = -\frac{r_{12} T^{*}_{1} \alpha_{12}}{K_{12}} + \frac{T^{*}_{1} w_{12} Z^{*}_{1} \beta_{1}}{(p_{12} + T^{*}_{1} + N^{*}_{1} \beta_{1})^2}$,
$a_{22} = -h_{12} + \frac{r_{12} (K_{12} - 2 T^{*}_{1} - N^{*}_{1} \alpha_{12})}{K_{12}} - \frac{w_{12} Z^{*}_{1} (p_{12} + N^{*}_{1} \beta_{1})}{(p_{12} + T^{*}_{1} + N^{*}_{1} \beta_{1})^2}$,
$a_{23} = -\frac{T^{*}_{1} w_{12}}{p_{12} + T^{*}_{1} + N^{*}_{1} \beta_{1}}$,
$a_{25} = h_{12}$,
$a_{31} = \frac{ Z^{*}_{1}c_{11} p_{11} w_{11}}{(N^{*}_{1} + p_{11})^2} + \frac{ Z^{*}_{1}c_{12} T^{*}_{1} w_{12} \beta_{1}}{(p_{12} + T^{*}_{1} + N^{*}_{1} \beta_{1})^2} $,
$a_{32} = -\frac{c_{12} w_{12} Z^{*}_{1} (p_{12} + N^{*}_{1} \beta_{1})}{(p_{12} + T^{*}_{1} + N^{*}_{1} \beta_{1})^2}$,
$a_{33} = -h_{13} - m_{1} + \frac{c_{11} N^{*}_{1} w_{11}}{N^{*}_{1} + p_{11}} - \frac{c_{12} T^{*}_{1} w_{12}}{p_{12} + T^{*}_{1} + N^{*}_{1} \beta_{1}}$,
$a_{36} = h_{13}$,
$a_{41} = h_{21}$,
$a_{44} = -h_{21} - \frac{p_{21} w_{21} Z^{*}_{2}}{(N^{*}_{2} + p_{21})^2} + \frac{r_{21} (K_{21} - 2 N^{*}_{2} - T^{*}_{2} \alpha_{21})}{K_{21}}$,
$a_{45} = -\frac{N^{*}_{2} r_{21} \alpha_{21}}{K_{21}}$,
$a_{46} = -\frac{N^{*}_{2} w_{21}}{N^{*}_{2} + p_{21}}$,
$a_{52} = h_{22}$,
$a_{54} = -\frac{r_{22} T^{*}_{2} \alpha_{22}}{K_{22}} + \frac{T^{*}_{2} w_{22} Z^{*}_{2} \beta_{2}}{(p_{22} + T^{*}_{2} + N^{*}_{2} \beta_{2})^2}$,
$a_{55} = -h_{22} + \frac{r_{22} (K_{22} - 2 T^{*}_{2} - N^{*}_{2} \alpha_{22})}{K_{22}} - \frac{w_{22} Z^{*}_{2} (p_{22} + N^{*}_{2} \beta_{2})}{(p_{22} + T^{*}_{2} + N^{*}_{2} \beta_{2})^2}$,
$a_{56} = -\frac{T^{*}_{2} w_{22}}{p_{22} + T^{*}_{2} + N^{*}_{2} \beta_{2}}$,
$a_{63} = h_{23}$,
$a_{64} =  \frac{Z^{*}_{2}c_{21} p_{21} w_{21}}{(N^{*}_{2} + p_{21})^2} + \frac{Z^{*}_{2}c_{22} T^{*}_{2} w_{22} \beta_{2}}{(p_{22} + T^{*}_{2} + N^{*}_{2} \beta_{2})^2} $,
$a_{65} = -\frac{c_{22} w_{22} Z^{*}_{2} (p_{22} + N^{*}_{2} \beta_{2})}{(p_{22} + T^{*}_{2} + N^{*}_{2} \beta_{2})^2}$,
$a_{66} = -h_{23} - m_{2} + \frac{c_{21} N^{*}_{2} w_{21}}{N^{*}_{2} + p_{21}} - \frac{c_{22} T^{*}_{2} w_{22}}{p_{22} + T^{*}_{2} + N^{*}_{2} \beta_{2}}$.

Following the theoretical analysis of Turing bifurcation in reaction-diffusion systems~\cite{XiaoChongZhang2012,LinXue2012,ToushengHuangCNSNS,FeifanZhang2022}, we introduced small perturbations to the heterogeneous steady state
\begin{equation}
	\left( \begin{matrix}
		N_1  \\
		T_1  \\
		Z_1  \\
		N_2  \\
		T_2  \\
		Z_2  \\		
	\end{matrix} \right)=\left( \begin{matrix}
		{{N}_{1}^{*}}  \\
		{{T}_{1}^{*}}  \\
		{{Z}_{1}^{*}}  \\
		{{N}_{2}^{*}}  \\
		{{T}_{2}^{*}}  \\
		{{Z}_{2}^{*}}  \\
	\end{matrix} \right)+\varepsilon\left( \begin{matrix}
		{{N }_{1k}}  \\
		{{T }_{1k}}  \\
		{{Z }_{1k}}  \\
		{{N }_{2k}}  \\
		{{T }_{2k}}  \\
		{{Z }_{2k}}  \\	
	\end{matrix} \right){{e}^{\lambda t+i\mathbf{k}\cdot \bm{\rho }}}+c.c.+O\left( {{\varepsilon }^{2}} \right),
	\label{perturbation EQ}
\end{equation}
where ${N }_{1k}$, ${T }_{1k}$, ${Z }_{1k}$, ${N }_{2k}$, ${T }_{2k}$, and ${Z }_{2k}$ are the perturbations corresponding to $N_1$, $T_1$, $Z_1$, $N_2$, $T_2$ and $Z_2$, respectively; $\bm{\rho }=(x,y)$ is the spatial vector in two dimensional space; $\lambda$ stands for the growth rate of heterogeneous perturbations in time; $i$ represents the imaginary unit and $i^2 = - 1$;  $\mathbf{k}$ is the corresponding vector of wavenumbers and $\mathbf{k}\cdot\mathbf{k}=k^2$; c.c. represents complex conjugate. The uniform steady state's linear instability ($\varepsilon \ll 1$) is derived from the dispersion relations. The characteristic equation for $\lambda_{1}$, $\lambda_{2}$, $\cdots$, $\lambda_{6}$ can be calculated from det$(J_k) = 0$, where
\begin{equation}
	J_{k}=J-J_{D}k^{2},
	\label{JacobianK}
\end{equation}
in which
\begin{equation}
	J_{D}=\begin{pmatrix}
		d_{11} & 0 & 0  & 0 & 0 & 0\\
		0 & d_{12} & 0  & 0 & 0 & 0\\
		-d_{31}Z^{*}_{1} & d_{32}Z^{*}_{1} & d_{13}  & 0 & 0 & 0\\
		0 & 0 & 0 &    d_{21} & 0 & 0\\
		0 & 0 & 0 &    0 & d_{22} & 0\\
		0 & 0 & 0 &    -d_{41}Z^{*}_{2} & d_{42}Z^{*}_{2} & d_{23}
	\end{pmatrix},
	\label{Jacobiand}
\end{equation}
From det$(J_k) = 0$, we get the eigenvalues as the roots of the following equation:
\begin{equation}
	\lambda^{6}+b_{5}(k^{2})\lambda^{5}+b_{4}(k^{2})\lambda^{4}+b_{3}(k^{2})\lambda^{3}+b_{2}(k^{2})\lambda^{2}+b_{1}(k^{2})\lambda+b_{0}(k^{2})=0,
	\label{JacobianeigenvaluesEQ}
\end{equation}

Based on the theory of Turing pattern formation, the sixth-order characteristic Eq.~\ref{JacobianeigenvaluesEQ} can be addressed through numerical computation to determine the eigenvalues $\lambda_k$. We choose $d_{31}$ as the control parameter for bifurcation analysis. According to the principles of Turing instability, the emergence of Turing patterns is initiated when
\begin{equation}
	\text{Re}(\lambda_k)=0, \text{Im}(\lambda_k)=0~\text{at} ~k=k_T \neq 0.
	\label{Turingcondition}
\end{equation}
Condition~(\ref{Turingcondition}) is satisfied only if the constant term of Eq.~(\ref{JacobianeigenvaluesEQ}) equals zero, i.e., $b_{0}(k^{2})=0$.

The characteristic equation~(\ref{JacobianeigenvaluesEQ}) derived above governs the linear stability of the coupled two-layer system. While its numerical solution yields the dispersion curves presented in the main text, further analytical insight can be gained by examining two limiting regimes. In what follows, ``weak'' and ``strong'' coupling refer to the ratio $2h/\delta_0$, where $\delta_0$ is the separation between the two layer modes at the dominant wavenumber; the crossover $2h\simeq\delta_0$ therefore lies at the synchronization threshold $h_c$ identified in Sec.~\ref{criticalcoupling}.

\section{Weak-coupling regime ($h \ll h_c$)}\label{weakcoupling}

Following the perturbative approach developed for multiplex networks~\cite{Asllani2014}, we treat the inter-layer coupling strength $h$ as a small parameter. Expanding the eigenvalues $\lambda_k = \lambda_k^{(0)} + h\lambda_k^{(1)} + \mathcal{O}(h^2)$ and applying Rayleigh-Schrödinger perturbation theory to the linearized operator $J_k = J_k^{(0)} + hV$, we obtain the first-order correction:

\begin{equation}
	\lambda_k^{(1)} = \frac{\langle v_k^{(0)} | V | v_k^{(0)} \rangle}{\langle v_k^{(0)} | v_k^{(0)} \rangle},
	\label{S9}
\end{equation}

where $V$ is the coupling matrix
\begin{equation}
	V = \begin{pmatrix}
		-\mathbb{I} & \mathbb{I} & 0 & 0 & 0 & 0 \\
		0 & -\mathbb{I} & 0 & 0 & \mathbb{I} & 0 \\
		0 & 0 & -\mathbb{I} & 0 & 0 & \mathbb{I} \\
		\mathbb{I} & 0 & 0 & -\mathbb{I} & 0 & 0 \\
		0 & \mathbb{I} & 0 & 0 & -\mathbb{I} & 0 \\
		0 & 0 & \mathbb{I} & 0 & 0 & -\mathbb{I}
	\end{pmatrix},
	\label{S10}
\end{equation}
where $v_k^{(0)}$ are the eigenvectors of the decoupled system (i.e., the eigenmodes of the block-diagonal matrix $J_k^{(0)}$). This correction quantifies how the two independent dispersion peaks (corresponding to the two layers) shift and approach each other as $h$ increases from zero. The merging of peaks observed in Fig.~1 of the main text occurs when the separation between the two peaks becomes comparable to their widths, a condition that can be expressed analytically as $|\lambda_{k_1}^{(0)} - \lambda_{k_2}^{(0)}| \sim \mathcal{O}(h)$.

\section{Strong-coupling regime ($h \gg h_c$)}\label{strongcoupling}

In the opposite limit, the two layers are tightly coupled and tend toward a synchronized state. Defining symmetric (in-phase) and antisymmetric (out-of-phase) modes as 
$U_s = (U_1+U_2)/\sqrt{2}$ and $U_a = (U_1-U_2)/\sqrt{2}$, where $U_i = (N_i, T_i, Z_i)^\top$, the linearized dynamics can be approximately block-diagonalized:

\begin{equation}
	\lambda_k^{(s)} \approx \lambda(\bar{J} - \bar{D}k^2), \qquad
	\lambda_k^{(a)} \approx \lambda(\bar{J} - \bar{D}k^2 - 2h),
	\label{S11}
\end{equation}
where $\bar{J} = (J_1+J_2)/2$ and $\bar{D} = (D_1+D_2)/2$ are the layer-averaged Jacobian and diffusion matrices. The antisymmetric mode decays at rate proportional to $2h$, while the symmetric mode persists. This explains why strong coupling enforces spatial synchronization across layers, consistent with the numerical observations in Figs.~2-3 of the main text. A similar block-diagonalization approach has been employed for two-layer activator-inhibitor systems with identical layer dynamics~\cite{CalderonBarreto2026}.

\section{Role of cross-diffusion}

Unlike the standard reaction-diffusion systems considered in the above references, our model includes cross-diffusion terms ($d_{31}Z_1\Delta N_1 - d_{32}Z_1\Delta T_1$, etc.) that render the diffusion matrix non-diagonal. This introduces additional complexity: the linearized diffusion matrix takes the form

\begin{equation}
	D = \begin{pmatrix}
		d_{11} & 0 & 0 & 0 & 0 & 0\\
		0 & d_{12} & 0 & 0 & 0 & 0\\
		-d_{31}Z_1^* & d_{32}Z_1^* & d_{13} & 0 & 0 & 0\\
		0 & 0 & 0 & d_{21} & 0 & 0\\
		0 & 0 & 0 & 0 & d_{22} & 0\\
		0 & 0 & 0 & -d_{41}Z_2^* & d_{42}Z_2^* & d_{23}
	\end{pmatrix}.
	\label{S12}
\end{equation}

The off-diagonal entries couple the stability of different species and modify the effective diffusion coefficients. In the weak coupling limit, these cross-terms can be treated as perturbations to the diagonal diffusion coefficients, yielding effective diffusivities $d_{i}^{\text{eff}} = d_i + \mathcal{O}(d_{3i}Z_i^*)$. This effective diffusion picture provides a mechanistic interpretation for the observed layer-specific pattern selection and the differential noise sensitivity across trophic levels reported in the main text.

\section{Critical coupling threshold}\label{criticalcoupling}

The Turing instability threshold is determined by the condition

\begin{equation}
	\max_k \operatorname{Re}[\lambda_k(h)] = 0.
	\label{S13}
\end{equation}

Solving this equation numerically via continuation in $h$ yields the critical coupling strength $h_c$ and the corresponding critical wavenumber $k_c$. This provides the theoretical basis for the synchronization transition reported in Fig.~4 of the main text. For the parameter sets used in our simulations, we find $h_c \approx 0.008$ to $0.015$, depending on the specific layer parameters, which is consistent with the numerical observation that synchronization emerges for $h > 0.01$.

The above analysis shows that the coupled system exhibits a bifurcation from two independent Turing modes to a single synchronized mode as $h$ crosses $h_c$. The synchronization is not merely a numerical observation but follows directly from the spectral properties of the linearized operator, with cross-diffusion terms adding a layer of biological realism that modifies the effective diffusion coefficients and thus the pattern selection rules.

\section{Linear response of the layer-difference mode to noise}\label{diffmodenoise}

The simulations of the main text show that inter-layer coupling both synchronizes the two layers and reduces their sensitivity to stochastic forcing. The two effects share a common origin, which can be made explicit within the linear-response approximation. Writing the perturbations in the symmetric-antisymmetric basis already used in Sec.~\ref{strongcoupling}, $u_s=(u_1+u_2)/\sqrt{2}$ and $u_a=(u_1-u_2)/\sqrt{2}$, the linearized dynamics of a Fourier mode with wavenumber $k$ takes the form
\begin{equation}
	\frac{d}{dt}\begin{pmatrix} u_s \\[0.4ex] u_a \end{pmatrix}
	=
	\begin{pmatrix}
		\bar{J}-\bar{D}k^{2} & \Delta J \\[0.8ex]
		\Delta J & \bar{J}-\bar{D}k^{2}-2h\,\mathbb{I}
	\end{pmatrix}
	\begin{pmatrix} u_s \\[0.4ex] u_a \end{pmatrix}
	+
	\begin{pmatrix} \xi_s \\[0.4ex] \xi_a \end{pmatrix},
	\label{S14}
\end{equation}
where $\bar{J}=(J_1+J_2)/2$, $\bar{D}=(D_1+D_2)/2$ and $\Delta J=(J_1-J_2)/2$ are the symmetric and antisymmetric parts of the Jacobian, $\mathbb{I}$ is the identity, and $\xi_s$ and $\xi_a$ are the symmetric and antisymmetric projections of the multiplicative noise. Because the two layers are forced independently, both projections have zero mean and variance $2\sigma^2A^2$ per unit time, with $A$ the amplitude of the unperturbed pattern. Two properties of Eq.~(\ref{S14}) are relevant. First, the antisymmetric block differs from the symmetric one only by the term $-2h\,\mathbb{I}$: the coupling adds a damping $2h$ that is independent of the wavenumber and acts on exactly the component that the synchronization errors $\delta N$, $\delta T$ and $\delta Z$ measure. Second, when the inter-layer mismatch is small compared with the coupling, $|\Delta J|\ll 2h$, the regime in which the two dispersion peaks have merged, the antisymmetric block decouples, and the growth rate of the antisymmetric mode at the dominant wavenumber is $\lambda_a=\delta_0-2h$, where $\delta_0\equiv\operatorname{Re}\lambda_a^{(0)}$ is the growth rate of the layer-difference mode in the absence of coupling. The antisymmetric mode is therefore stable only when
\begin{equation}
	2h>\delta_0, \qquad \text{i.e.}\qquad h>h_c=\delta_0/2,
	\label{S15}
\end{equation}
which identifies $h_c$ with half the mismatch between the two layer modes and is consistent both with the peak-merging condition of Secs.~\ref{weakcoupling} and~\ref{strongcoupling} and with the numerical estimate of $h_c$ obtained in Sec.~\ref{criticalcoupling}.

For $h>h_c$ the antisymmetric component obeys a damped Ornstein-Uhlenbeck equation with damping rate $2h-\delta_0$, so that its stationary variance is $\langle u_a^2\rangle=\sigma^2A^2(2h-\delta_0)^{-1}$ and its stationary amplitude is
\begin{equation}
	\langle \delta X\rangle \simeq \sigma A\,(2h-\delta_0)^{-1/2}, \qquad h>h_c,
	\label{S16}
\end{equation}
up to a constant of order unity. Two remarks are in order. First, the factor $(2h-\delta_0)^{-1/2}$ acts on the difference between the two layers, so the noise-driven part of the synchronization error is suppressed as $h$ grows; this is consistent with the monotonic decrease of the measured steady-state error with $h$ in Fig.~4(b) of the main text. Second, Eq.~(\ref{S16}) describes only that noise-driven increment. In our simulations the measured error at weak coupling is dominated by the deterministic mismatch between the layers that survives at $\sigma=0$, and noise can even lower the total error, because it degrades the pattern amplitude and with it the mismatch itself. Equation~(\ref{S16}) should therefore not be read as a prediction of the total error, and it does not give the noise intensity at which the pattern collapses.

A further implication of Eq.~(\ref{S16}) is that the noise level at which the layer difference becomes comparable to the pattern amplitude should be pushed to larger values as $h$ increases. We do not test this prediction here, because the pattern amplitude under strong forcing is set by the nonlinear terms that the linear analysis discards.

Equation~(\ref{S16}) is a linear-response estimate. It neglects the nonlinear saturation of the pattern amplitude, the dependence of $A$ on $h$, and the residual coupling between the symmetric and antisymmetric blocks retained through $\Delta J$ in Eq.~(\ref{S14}), so it should be read as a scaling law for how the layer difference responds to stochastic forcing rather than as a quantitative prediction. Going beyond it requires the amplitude equations of the coupled system, which we leave for future work.


\clearpage
\section{Supplemental Tables}
\begin{table}[h]
	\centering
	\caption{ Parameter values used in the simulations of Figs. 1-7 of the main text and of Figs. S3 and S6.}
	\resizebox{\textwidth}{!}{ 
		\begin{tabular}{cccccccccccccccccccccccc}
			\toprule
			{Figure} &  {Layer} & \multicolumn{21}{c}{Parameter values} \\
			\cmidrule(lr){3-23}
			& & $r_{i1}$ & $r_{i2}$ & $\alpha_{i1}$ & $\alpha_{i2}$ & $K_{i1}$ & $K_{i2}$& $c_{i1}$ & $c_{i2}$ & $w_{i1}$ & $w_{i2}$ & $p_{i1}$ & $p_{i2}$ & $\beta_{i}$ & $m_{i}$&$d_{i1}$ & $d_{i2}$ & $d_{i3}$& $d_{3i}$ & $d_{4i}$& $h_{il}$ & $\sigma$\\
			\midrule
			{Figs.~1,~2a,~3a,~4,~5,~7a,~7d} & 1&0.56 & 0.49 & 0.25 & 0.1 & 1600 & 1500 & 0.45 & 0.4 & 0.45 & 0.4 & 150 & 140& 2 & 0.1 & 1 & 0.1 & 0.001 & 0.0006 & 0.0005 & [0,0.06] & [0,0.06]   \\
			& 2 & 0.56 & 0.49 & 0.25 & 0.1 & 1600 & 1500 & 0.45 & 0.4 & 0.45 & 0.4 & 150 & 140& 2 & 0.1 & 1 & 0.1 & 0.001 & 0.0012 & 0.0007 & [0,0.06] & [0,0.06]\\
			\addlinespace
			{Figs.~2b,~3b,~6} &  1 &0.56 & 0.49 & 0.25 & 0.1 & 800 & 800 & 0.45 & 0.4 & 0.45 & 0.4 & 150 & 140& 2 & 0.1 & 1 & 0.1 & 0.001 & 0.0015 & 0.001 & [0,0.06] & [0,0.06]\\
			&  2 & 0.56 & 0.49 & 0.25 & 0.1 & 800 & 800 & 0.45 & 0.4 & 0.45 & 0.4 & 150 & 140& 2 & 0.1 & 1 & 0.1 & 0.001 & 0.0009 & 0.0003 & [0,0.06] & [0,0.06]\\
			\addlinespace
			{Figs.~2c,~3c,~7b,~7e} &  1 &0.56 & 0.49 & 0.25 & 0.1 & 1500 & 1500 & 0.45 & 0.4 & 0.5 & 0.5 & 80 & 70& 2 & 0.1 & 1 & 0.1 & 0.001 & 0.0004 & 0.0002 & [0,0.06] & [0,0.06]\\
			&  2 & 0.56 & 0.49 & 0.25 & 0.1 & 1500 & 1500 & 0.45 & 0.4 & 0.5 & 0.5 & 80 & 70& 2 & 0.1 & 1 & 0.1 & 0.001 & 0.0012 & 0.0008 & [0,0.06] & [0,0.06]\\
			\addlinespace
			{Figs.~2d,~3d,~7c,~7f} &  1 &0.66 & 0.62 & 0.25 & 0.1 & 1200 & 1200 & 0.45 & 0.4 & 0.5 & 0.5 & 50 & 30& 2 & 0.1 & 1 & 0.1 & 0.001 & 0.001 & 0.0005 & [0,0.06] & [0,0.06]\\
			& 2 &0.56 & 0.59 & 0.25 & 0.1 & 1200 & 1200 & 0.45 & 0.4 & 0.5 & 0.5 & 50 & 30& 2 & 0.1 & 1 & 0.1 & 0.001 & 0.0012 & 0.0008 & [0,0.06] & [0,0.06] \\	
			\addlinespace
			{Fig.~S3} & 1 &0.6 & 0.5 & 0.25 & 0.1 & 1500 & 1500 & 0.45 & 0.45 & 0.5 & 0.5 & 170 & 150 & 2.1 & 0.1 & 1 & 0.1 & 0.0001 & $1.2d_{32}$ & $[0.5,3]\times10^{-3}$ & [0,0.1] & 0\\
			& 2 &0.6 & 0.5 & 0.25 & 0.1 & 1500 & 1500 & 0.45 & 0.45 & 0.5 & 0.5 & 170 & 150 & 2.1 & 0.1 & 1 & 0.1 & 0.0001 & 0.0026 & 0.0027 & [0,0.1] & 0\\	
			\bottomrule
	\end{tabular}}
	\label{tab1}	
\end{table}

\clearpage
\section{Supplemental Figures}

\begin{figure}[H]
	\centering
	\includegraphics[width=17cm]{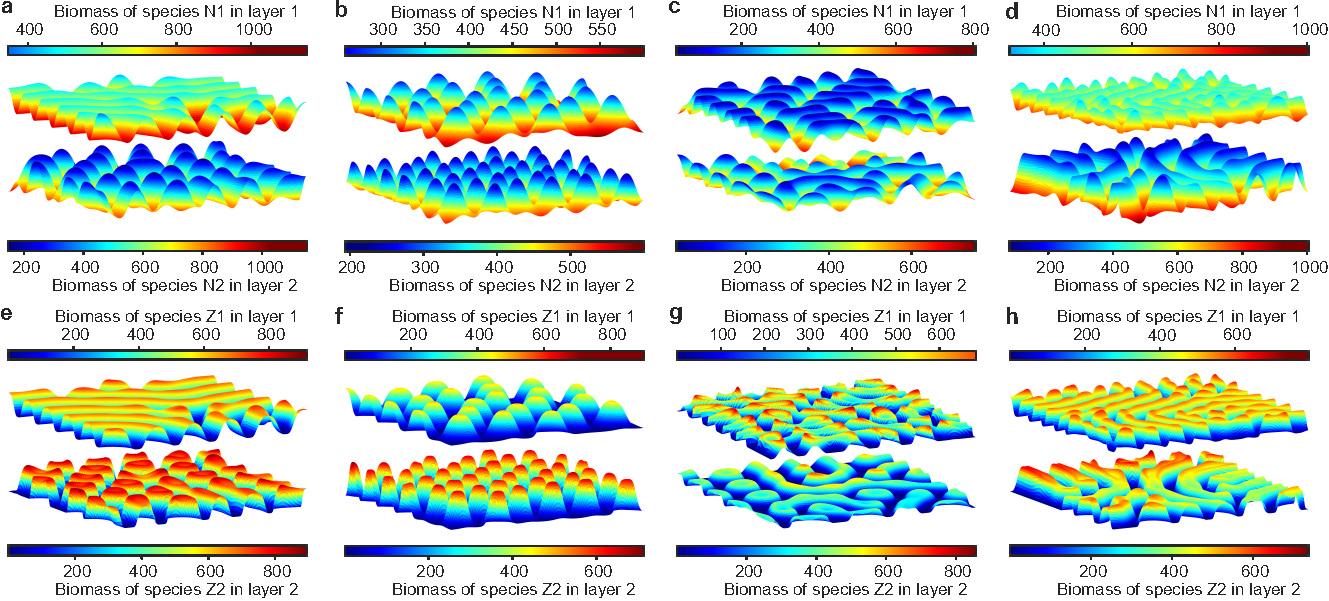}
	\caption{\label{3DpatternNonSychSM} Self-organized pattern dynamics in the non-coupled aquatic system (without vertical migration). (a-h) Spatial patterns in the surface (Layer 1) and bottom (Layer 2) layers exhibit distinct structural forms, such as spots, stripes, and mixtures. The parameter values are the same as Fig.~2.}	
\end{figure}

\begin{figure}[H]
	\centering
	\includegraphics[width=17cm]{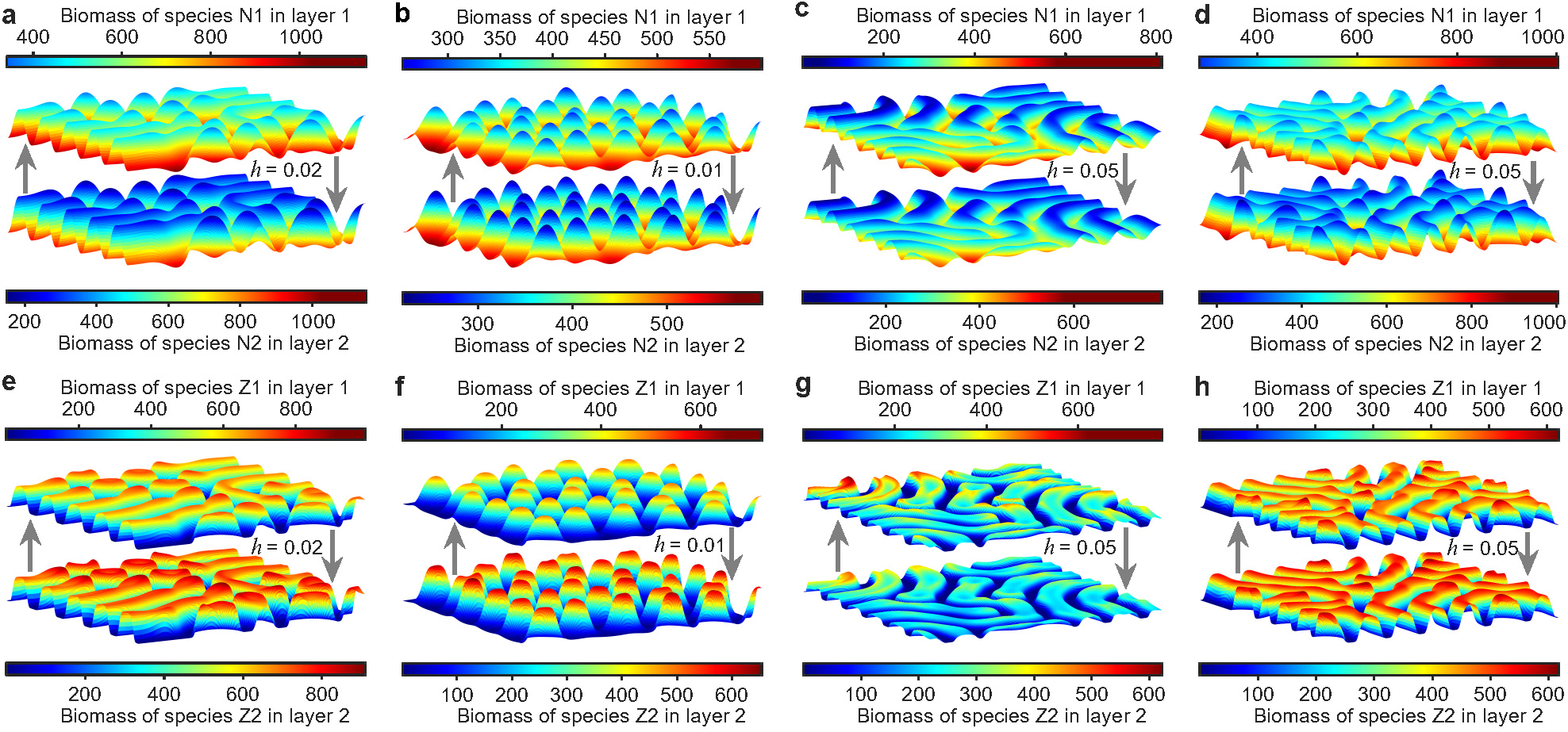}
	\caption{\label{3DpatternSychSM} Self-organized synchronization pattern dynamics in the coupled aquatic system with vertical migration. (a-h) With the introduction of coupling, the initially independent spatial patterns from Fig.~3 (e.g., spots, stripes, mixtures) evolve into synchronized states across layers. The parameter values are the same as Fig.~3.}	
\end{figure}

\begin{figure}[H]
	\centering
	\includegraphics[width=17cm]{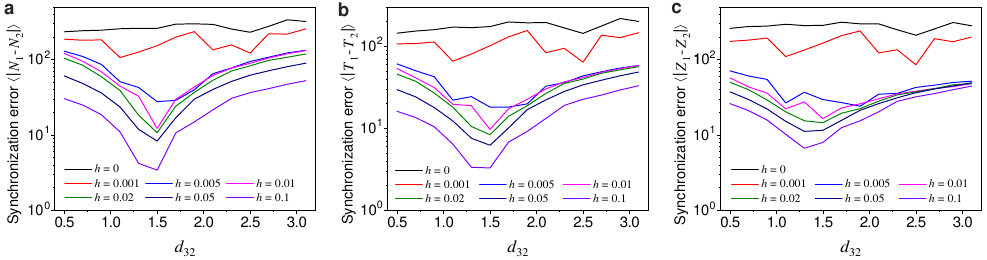}
	\caption{\label{FigS6}
		Synchronization error as a function of the cross-diffusion coefficient $d_{32}$ for several interlayer coupling strengths $h$: (a) $\langle|N_{1}-N_{2}|\rangle$, (b) $\langle|T_{1}-T_{2}|\rangle$ and (c) $\langle|Z_{1}-Z_{2}|\rangle$, on a logarithmic vertical scale. Layer~1 uses $d_{31}=1.2\,d_{32}$ and layer~2 uses $d_{41}=0.0026$ and $d_{42}=0.0027$; all other parameter values are listed in the last row of Table~\ref{tab1}, which differs from the four parameter sets of the main text in its growth rates, grazing rates and cross-diffusion coefficients. At every $d_{32}$ the error decreases systematically with $h$, the steepest decrease occurring around the synchronization threshold $h\simeq0.01$ to $0.02$, and at $h=0.1$ the error is one to two orders of magnitude smaller than in the uncoupled case. The error reaches a minimum near $d_{32}\simeq1.2\times10^{-3}$ and increases again for larger $d_{32}$. The same qualitative dependence on $h$ and $d_{32}$ is found for the three species. Simulations are deterministic ($\sigma=0$) on a $201\times201$ grid, run for $10^{6}$ time steps at $\delta t=0.01$. The error is averaged over the domain and over the final state.}
\end{figure}

\begin{figure}[H]
	\centering
	\includegraphics[width=17cm]{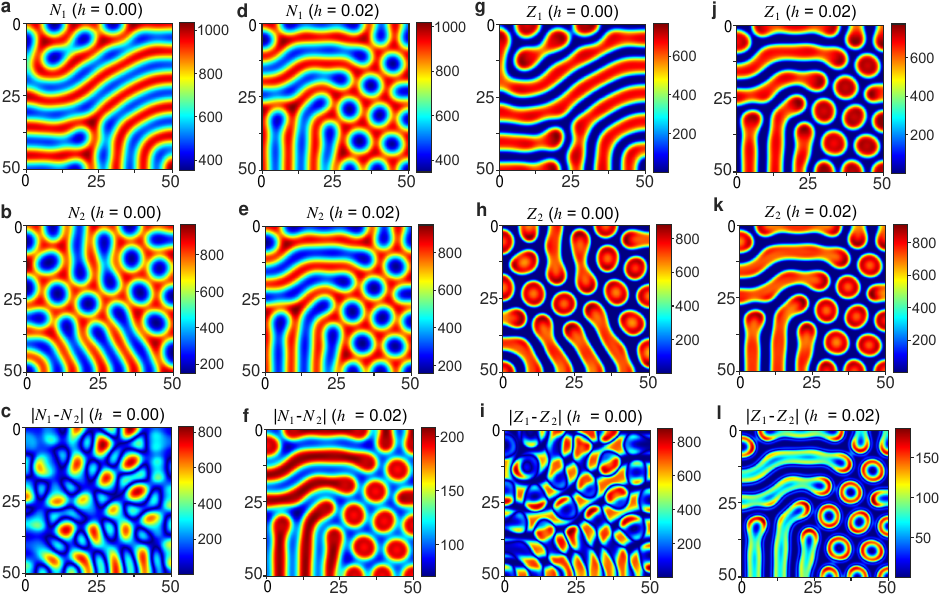}
	\caption{\label{sync_error_FigSM} 
		Visual comparison of self-organized patterns and their synchronization errors for non-coupled (Fig.~\ref{sync_error_FigSM}) and coupled (Fig.~\ref{sync_error_FigSM}) systems. (a-l) For the non-coupled system ($h = 0$): (a-b, g-h) self-organized patterns, (c, i) pattern of the synchronization error, which exhibits a structure distinctly different from (a-b, g-h).  (d-e, j-k) For the coupled system ($h = 0.02$): (d-e) synchronized patterns, (f, l) pattern of the synchronization error, which maintains structural similarity to (d-e, j-k). The parameter values are the same as Fig.~5.} 	
\end{figure}

\begin{figure}[H]
	\centering
	\includegraphics[width=17cm]{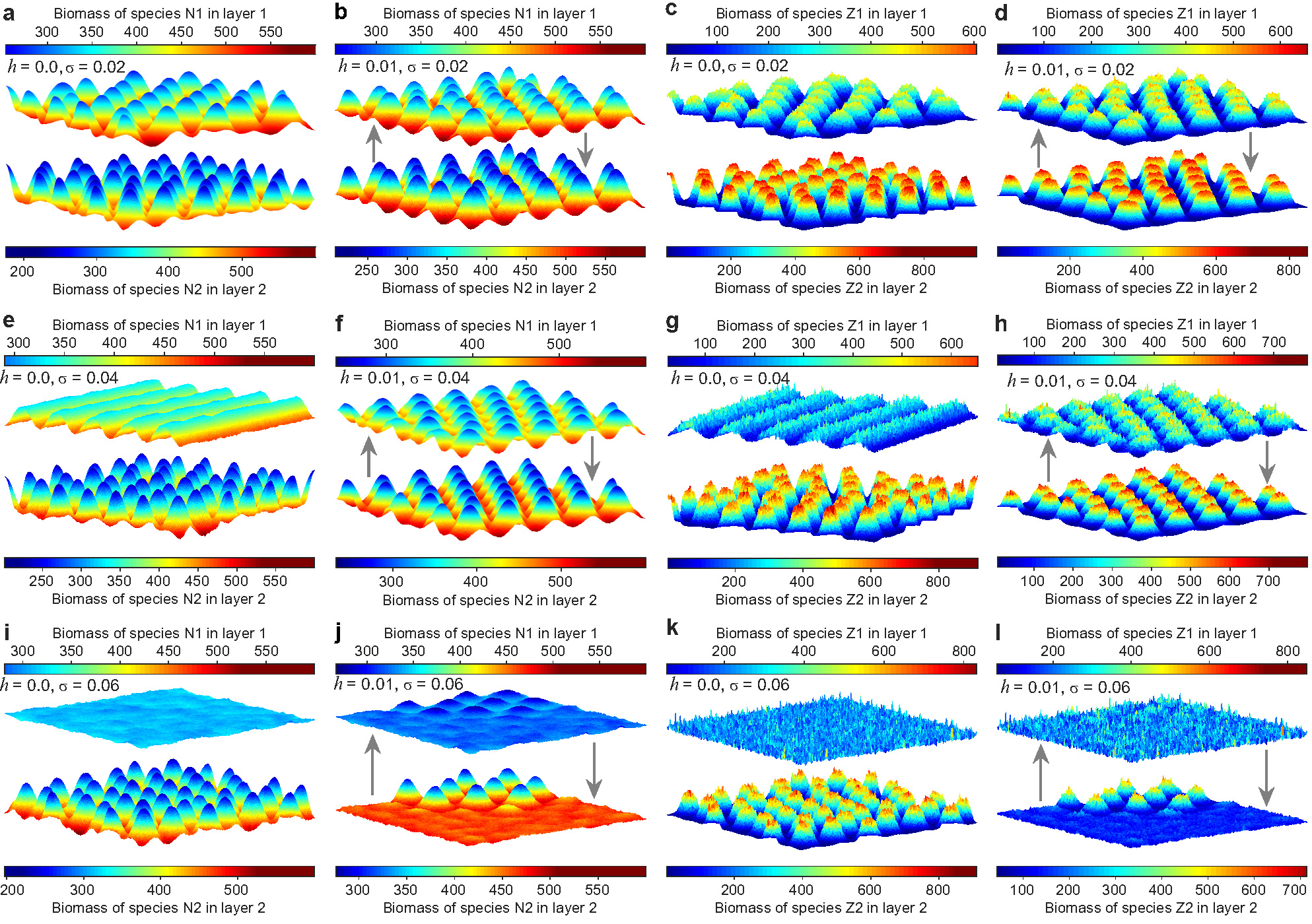}
	\caption{\label{FigdotSM} 
		Effect of stochasticity on spotted pattern structures. (a, c, e, g, i, k) Without coupling, increasing noise intensity affects the upper layer more than the lower layer: the upper-layer pattern transitions from spots to stripes and eventually vanishes, while the lower layer maintains its spotted structure. (b, d, f, h, j, l) With coupling, the system exhibits robustness to stochastic fluctuations, preserving the spotted structure in both layers until the noise intensity exceeds a critical threshold, beyond which the spatial patterns collapse.
		The parameter values are the same as Fig.~6.} 	
\end{figure}

\begin{figure}[H]
	\centering
	\includegraphics[width=17cm]{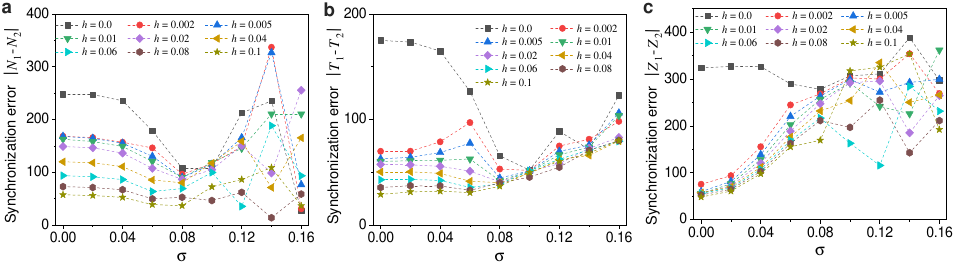}
	\caption{\label{FigS7}
		Synchronization error as a function of the noise intensity $\sigma$ for interlayer coupling strengths $h$ between $0$ and $0.1$: (a) $\langle|N_{1}-N_{2}|\rangle$, (b) $\langle|T_{1}-T_{2}|\rangle$ and (c) $\langle|Z_{1}-Z_{2}|\rangle$. The parameter values are those of the first row of Table~\ref{tab1}. The simulations use a $201\times201$ grid with $\delta t=0.01$ and $10^{5}$ time steps, and each point is an average over three realizations of the noise that start from the same random initial field. For $\sigma\le0.06$ the error decreases monotonically with $h$ for all three species. Beyond $\sigma\simeq0.1$ the patterns are strongly degraded, and the errors no longer vary monotonically, neither with $\sigma$ nor with $h$. Between $\sigma=0$ and $\sigma=0.08$, $\langle|Z_{1}-Z_{2}|\rangle$ increases by a factor of three to four for every $h>0$, whereas $\langle|N_{1}-N_{2}|\rangle$ decreases at every $h$ and $\langle|T_{1}-T_{2}|\rangle$ decreases for $h\le0.06$.}
\end{figure}

\begin{figure}[H]
	\centering
	\includegraphics[width=17cm]{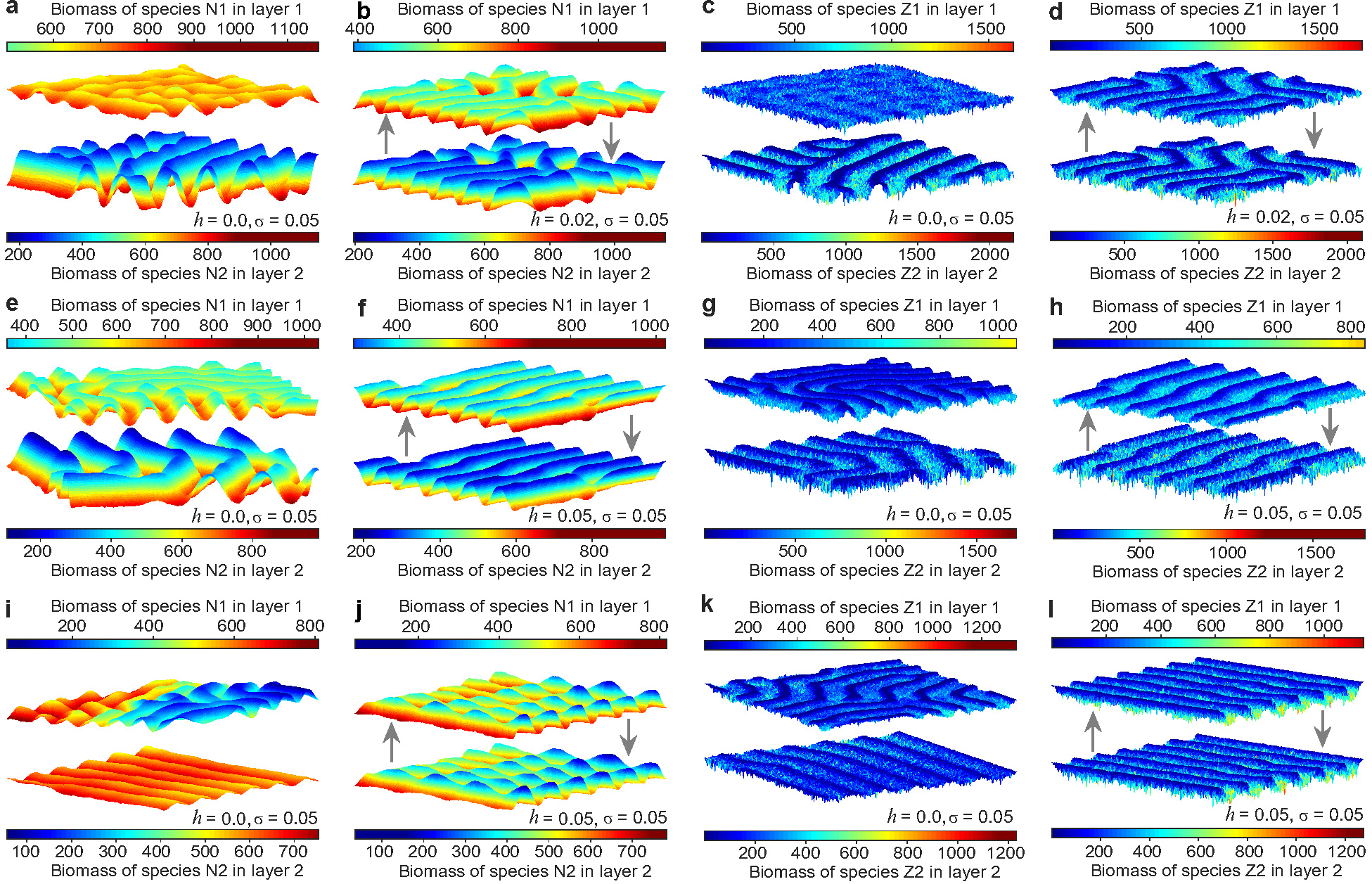}
	\caption{\label{FignoiseSM} 
		Robust synchronization patterns by coupling in a stochastic environment. (a, c, e, g, i, k) Without coupling, the two layers maintain distinct and independent spatial patterns in the presence of noise. (b, d, f, h, j, l) With coupling introduced, the initially different patterns achieve a robust synchronized state, demonstrating that coupling can overcome both initial heterogeneity and stochastic perturbations to establish synchronous dynamics.
		The parameter values are the same as Fig.~7.} 	
\end{figure}

\end{document}